\documentclass[amsmath,nobibnotes,aps,amssymb,pra, prl,aps,showpacs,superscriptaddress,twocolumn, longbibliography, reprint]{revtex4-2}
\usepackage{color}
\usepackage{graphicx}
\usepackage{tikz}
\usepackage{epsfig,psfrag}
\usepackage{amsmath}
\usepackage{amssymb}
\usepackage{amsbsy}
\usepackage{amsthm}
\usepackage{amsfonts}
\usepackage{wasysym}
\usepackage{bbm}
\usepackage{tabularx}
\usepackage{euscript}
\usepackage{enumerate}
\usepackage{amsfonts}
\usepackage{exscale}
\usepackage{bbold}
\usepackage{float}
\usepackage{slashed}
\usepackage{caption}
\usepackage{subcaption}
\usepackage{comment}
\usepackage[colorlinks,citecolor=blue]{hyperref}
\usepackage{ragged2e}

\usepackage[title]{appendix}

\usepackage{mathrsfs}  

\usepackage{xr}
\usepackage{scalerel,stackengine}
\stackMath
\newcommand\reallywidehat[1]{%
	\savestack{\tmpbox}{\stretchto{%
			\scaleto{%
				\scalerel*[\widthof{\ensuremath{#1}}]{\kern-.6pt\bigwedge\kern-.6pt}%
				{\rule[-\textheight/2]{1ex}{\textheight}}%WIDTH-LIMITED BIG WEDGE
			}{\textheight}% 
		}{0.5ex}}%
	\stackon[1pt]{#1}{\tmpbox}%
}
\newcommand{\bea}{\begin{eqnarray}}
	\newcommand{\eea}{\end{eqnarray}}
\newcommand{\ba}{\begin{array}}
	\newcommand{\ea}{\end{array}}

\usepackage[normalem]{ulem}

\def\bea{\begin{eqnarray}}
	\def\eea{\end{eqnarray}}
\def\Tr{\mathrm{Tr}}
\def\nn{\nonumber}
\def\bea{\begin{eqnarray}}
	\def\eea{\end{eqnarray}}
\def\nn{\nonumber}
\def\ba{\begin{array}}
	\def\ea{\end{array}}
\def\nn{\nonumber}
\def\Tr{\text{Tr}}

\def\red{\textcolor[rgb]{0.98,0.00,0.00}}

\def\newred{\textcolor[rgb]{0.98,0.00,0.00}}

\newcommand{\dd}{\text{d}}
\newcommand{\e}{\text{e}}

\renewcommand{\red}[1]{#1}        % Just prints the text normally
\renewcommand{\sout}[1]{}         % Swallows the argument completely

\renewcommand{\newred}[1]{#1}        % Just prints the text normally

\begin{document}
	
	\title{Many-body spectral transitions through the lens of the variable-range SYK$_2$ model}
	%\title{Many-body spectral transitions in the variable-range SYK$_2$ model}

	\author{Andrea Legramandi}
	\email{andrea.legramandi@unitn.it}
	\affiliation{Pitaevskii BEC Center, CNR-INO and Dipartimento di Fisica, Universit\'a di Trento, I38123 Trento, Italy}
	\affiliation{INFN-TIFPA, Trento Institute for Fundamental Physics and Applications, Trento, Italy}

	\author{Soumik Bandyopadhyay}
	\email{soumik.bandyopadhyay@unitn.it}
	\affiliation{Pitaevskii BEC Center, CNR-INO and Dipartimento di Fisica, Universit\'a di Trento, I38123 Trento, Italy}
	\affiliation{INFN-TIFPA, Trento Institute for Fundamental Physics and Applications, Trento, Italy}

	\author{Philipp Hauke}
	\email{philipp.hauke@unitn.it}
	\affiliation{Pitaevskii BEC Center, CNR-INO and Dipartimento di Fisica, Universit\'a di Trento, I38123 Trento, Italy}
	\affiliation{INFN-TIFPA, Trento Institute for Fundamental Physics and Applications, Trento, Italy}

	\begin{abstract}
		The Sachdev–Ye–Kitaev (SYK) model is a cornerstone in the study of quantum chaos and holographic quantum matter. Real-world implementations, however, deviate from the idealized all-to-all connectivity, raising questions about the robustness of its chaotic properties. In this work, we investigate a quadratic SYK model with distance-dependent interactions governed by a power-law decay. By analytically and numerically studying the spectral form factor (SFF), we uncover how  transitions present in the single-particle limit carry over to the many-body system. 
		Non-trivial cancellations in the one-loop contributions lead to a robustness of the SFF under a considerable reduction of the interaction range. Further suppression leads to a breakdown of perturbation theory around the infinite-range path-integral saddle and the appearance of new spectral regimes, marked by a higher dip and the emergence of a secondary plateau. Our results highlight the interplay between single-particle criticality and many-body dynamics, offering new insights into the quantum chaos-to-localization transition and its reflection in spectral statistics.
	\end{abstract}

	\maketitle 
	
	{\it Introduction.} The Sachdev--Ye--Kitaev (SYK) model ~\cite{Kitaev_talks,SachdevYe_1993} has emerged as a paradigmatic example in the study of quantum chaos and holographic quantum matter~\cite{Sachdev_2015,Jensen_2016, book_holoquantmat, Chowdhury_etal2022}, thanks to its analytical tractability in the thermodynamic limit~\cite{MaldacenaStanford_2016}. Additionally, it has inspired various proposals for experimental implementation~\cite{Danshita_etal2017,PikulinFranz_2017,GarciaAlvarez_etal2017,Chew_etal2017,Babbush_etal2019,Chen_etal2018,Luo_etal2019,WeiSedrakyan_2021,Uhrich_etal2023,Baumgartner_2024}, which could open the door to studying the theoretically challenging regime of intermediate system sizes, where quantum corrections to gravity become prominent.
	However, any realistic laboratory implementation would deviate from the idealized SYK model, which assumes all-to-all connectivity and perfectly uncorrelated random couplings. These deviations raise important questions about the robustness of the model's properties when the interactions are constrained to a finite range. The introduction of distance-dependent interactions, as is typical for platforms such as trapped ions~\cite{Britton_2012,Jurcevic_2014,Smith_etal2016,Trautmann_2018}, cold atoms and molecules~\cite{Chomaz_2023,BrowaeysLahaye_2020,Cornish_2024}, solid state systems~\cite{Awschalom_2018}, etc., is expected to considerably impact the system’s chaotic behavior~\cite{Chen_2019} and can potentially lead to significant many-body phenomena such as non-ergodicity and localization \cite{Burin_2015_1,Burin_2015_2,Hauke_2015_MBL,Tikhonov_2018, Deng_2018,Thomson_2020,Deng_2019,Nag_2019,Roy_2019,Deng_2020,Modak_2020_1,Modak_2020_2,Cheng_2023}, which are central to understanding the quantum dynamics of strongly coupled but not chaotic systems.
	
	\begin{figure}[h!]
		\includegraphics[width=0.42\textwidth]{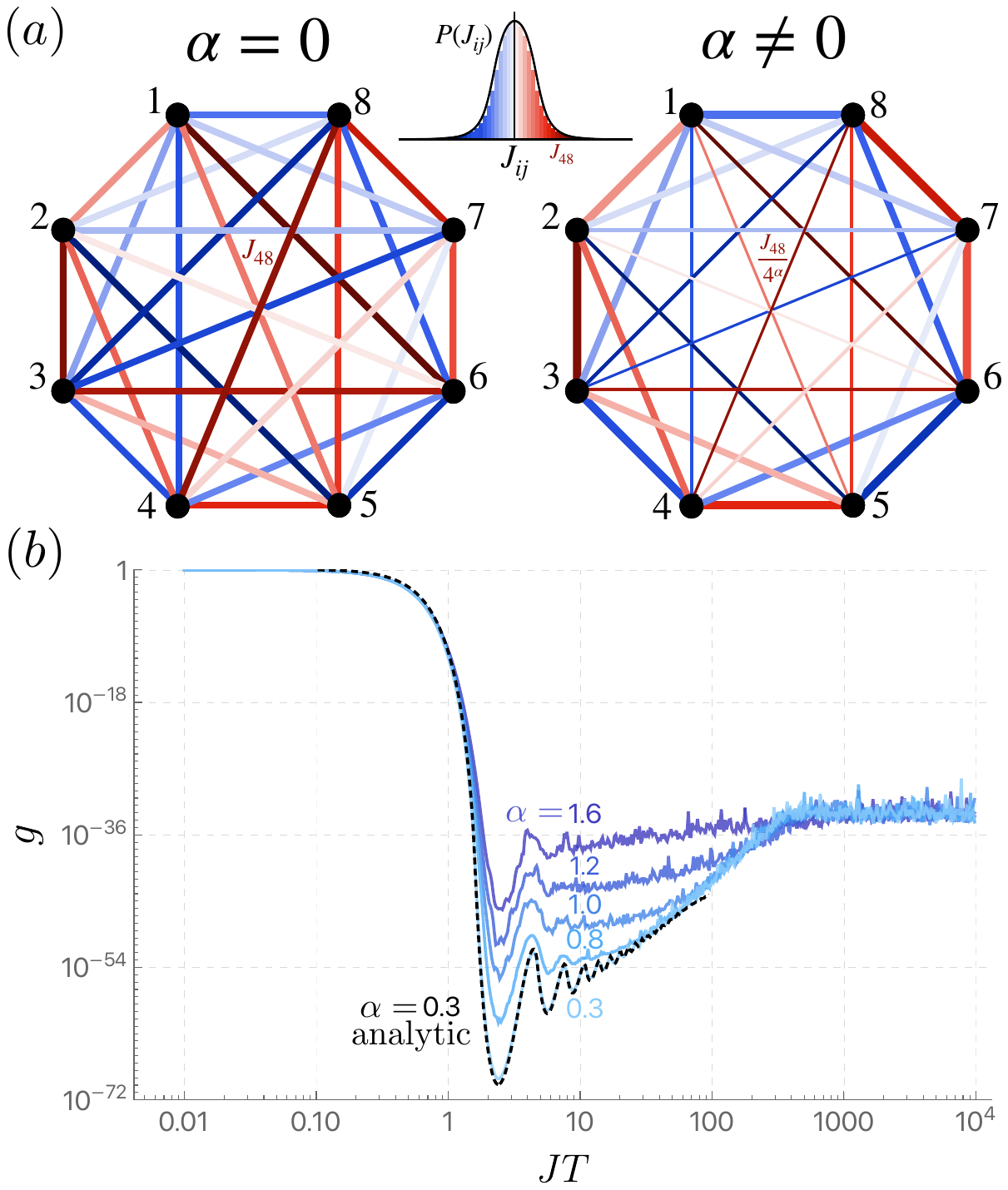}
		\vspace{-1.8cm} \caption{\justifying
			(a) Illustration of the  variable-range SYK$_2$-model given in Eq.~\eqref{eq:model_action} for $N=8$ Majorana fermions on a circle, with all-to-all random couplings drawn from a Gaussian distribution. The line thickness represents the interaction strength, which decays with distance as $r^{-\alpha}$.
			(b) The SFF $g$ for $N=200$ fermions, 
			%shown for $\alpha=0.3, 0.8, 1.0, 1.2, 2.5$ (from light to dark blue shades).
			for increasing $\alpha$ from light blue to violet. 
			For $\alpha < 1/2$, the numerical results align well with the analytical prediction from Eq.~\eqref{eq:g_analy} (black dashed line). For $\alpha > 1/2$, the system moves beyond the validity regime of the analytical approach, indicating a drastic change of the physics, as also suggested by
			a higher dip and the emergence of a secondary plateau, which merges with the late-time plateau for $\alpha \gtrsim 3/2$.
		}
		\label{fig:SFF}
	\end{figure}

	In this work, we introduce and study the transitions of a system composed of $N$ Majorana fermions with random hopping suppressed by a power-law decay $r^{-\alpha}$ (see Fig.~\ref{fig:SFF}a)---a version of the quadratic SYK that in the limit of a single particle becomes a power-law random banded matrix (PRBM) model \cite{Mirlin_etal1996}. 
	The PRBM has become a popular model for studying Anderson criticality \cite{Anderson_1958} and multifractality \cite{janssen1994, EversMirlin_2008}, as it is amenable to analytical studies using the supersymmetric sigma-model approach \cite{Fyodorov_Mirlin1991}. %, from which several quantities can be derived, such as the inverse participation ratio and the corrections to the energy spectrum.
	Here, we take a different route: we study the many-body spectral statistics of the SYK model using the average spectral form factor (SFF) \cite{Brezin_Hikami1997}. In the thermodynamic limit, it can be treated analytically thanks to a path integration \cite{Saad_etal2018}, which results in a different sigma model than the supersymmetric one.
	
	The SFF, first introduced in the context of random matrix theory \cite{Brezin_Hikami1997}, reflects the distribution of energy levels in a system. In chaotic systems, the average SFF typically exhibits a universal structure characterized by an initial non-universal slope, a dip, a linear ramp, and a late-time plateau. However, in non-chaotic systems, the average SFF loses this universal behavior and is less well understood \cite{Kos_etal2017,Gharibyan_etal2018,Chan_2019,Suntajs_etal2019,nivedita_etal2020,prakash_2020,Prakash_2021,Suntajs_2021,Nandy_etal2022,Anegawa_etal2023,barney_etal2023,Hopjan_Lev2023,Baumgartner_etal2024,Orman_Preskill2024,Buijsman_2024,dong_2024}. For example, in the quadratic SYK model, it displays an exponential ramp instead of a linear one \cite{Winer:2020mdc,Liao_etal2020}, highlighting that the system is chaotic only at the single-particle level (see also \cite{Flynn_etal2024}).
	
	Our model allows us to take this analysis further by investigating how a variable interaction range impacts SYK physics. 
		A central question is to what extent transitions known in the single-particle limit carry over to the many-body system, whose behavior can differ significantly due to fermionic statistics \cite{kruger2008fermionic,Barnum:2004zz} and single-particle mobility edges \cite{sodin2010spectral}. In particular, we aim to determine whether the SFF can serve as a reliable marker for such transitions.
As a saddle point calculation reveals, an initial decrease of the interaction range affects the SFF only subleadingly. This robustness derives from a non-trivial cancellation of one-loop contributions, demonstrating that the theory preserves ergodicity over an extended parameter range. 
		However, once the interaction range is decreased beyond $\alpha\sim 1/2$,
	perturbation theory breaks down and a transition occurs marked by distinctive changes in the SFF (Fig.~\ref{fig:SFF}b): a growth of the dip and the emergence of a secondary plateau, indicative of a pre-thermalization regime \cite{Gharibyan_etal2018,Hopjan_Lev2023} where the Hilbert space is not fully explored \cite{Baumgartner_etal2024}. The secondary-plateau height increases with $\alpha$, exhibiting its steepest growth near $\alpha \sim 1$, until it reaches the late-time plateau at $\alpha \sim 3/2$, echoing the single-particle PRBM transitions.
These features allow us to identify and characterize many-body level-statistics transitions, reveal their connection to single-particle localization transitions, and provide a robust diagnostic extendable to interacting systems.

	{\it Model.} We introduce the variable-range SYK$_2$ model as a variant of the SYK$_2$ Hamiltonian where the random two-fermion term is weighted by a function of the distance between the two sites involved,  
	\begin{equation}
		\label{eq:model_action}
		\begin{split}
			\mathcal{H} =& \frac{i}{\mathcal{N}} \sum_{i<j}^N J_{ij }a(i-j) \psi_i \psi_j \, ,\\
			a(i-j) =& (\min \left\{|i-j|,N-|i-j|\right\})^{-\alpha} \, .
		\end{split}
	\end{equation}
	The couplings $J_{ij}$ are real Gaussian variables with variance $\langle J_{ij} J_{kl} \rangle = J^2 \delta_{ik} \delta_{jl}$ and the Majorana fermions are distributed on a line with periodic boundary conditions. 
	For later convenience, we define the circular matrix $A$ 
	\begin{equation}
		\label{eq:A_def}
		A_{ij} = \left[a(i-j)\right]^2 / \mathcal{N}^2 \, , \qquad A_{ii}=0 \, ,
	\end{equation} 
	with $\mathcal{N}^2=\sum_{i} [a(i)]^2$. 
	
	The one-particle Hamiltonian $J_{ij}a(i-j)$, i.e., the coupling constant matrix, is an example of a PRBM model. It determines the excitations of the system in the single-particle sector \footnote{The notion of single-particle Hamiltonian might appear ambiguous for Majorana fermions. However, $N$ Majorana modes can be mapped to $N/2$ complex ones, allowing the Hamiltonian to be diagonalized via a Bogoliubov transformation. For number-conserving complex-fermion Hamiltonians, one can add a chemical potential to force the system to favor certain particle sectors. This approach can be used in the context of the path-integral calculations developed in the next sections. In this formulation, the single-particle limit can be recovered by taking the chemical potential to minus infinity.} and  exhibits different properties depending on the value of $\alpha$ \cite{Mirlin_etal1996}: for $\alpha \in [0,0.5)$, the PRBM is ergodic and indistinguishable from the Gaussian orthogonal ensemble. For $\alpha> 1.5, $ it becomes integrable \cite{Xu_2023}. At $\alpha = 1$, it has an Anderson critical point \cite{Anderson_1958,Mirlin_etal1996,Kravtsov_1997,Evers_2000,Varga_2000,Mirlin_2000} separating the delocalized but non-ergodic regime $\alpha \in (0.5,1)$~\cite{Bogomolny_2018,Nosov_2019,Xu_2023}, from the localized but super-diffusive regime $\alpha \in (1,1.5)$. In this paper, we explore the transitions of the corresponding many-body system, the variable-range SYK$_2$ model.

	{\it Path integral for the average SFF.} The main quantity we will use to analyze the many-body physics is the disorder-averaged SFF, defined as $g(T)= {\langle |Z(i T)|^2 \rangle}/{|Z(0)|^2}$.
	The angle brackets denote averaging over the random couplings, while $Z(0)$ is the Hilbert-space dimension. 
	%\red{ (which for $|Z(0)|^2 = L^2$ is trivial).}
	In what follows, we express $g(t)$ using collective-field variables, determine the dominant saddle-point solutions that spontaneously break some action symmetries, and identify the zero-mode contributions, which depend on the eigenvalues of $A$ defined in Eq.~\eqref{eq:A_def}.

	First, we write the partition function as a path integral,
	\begin{equation}
		Z(i T) \red{=\!\text{Tr} U\! }= \! \int \! \mathcal{D} \psi \exp \bigg\{\int_0^T \Big(i \mathcal{H}- \! \sum_i \frac12 \psi_i \partial_t \psi_i \Big) \dd t \bigg\} ,
	\end{equation}
	with $U = e^{-i T \mathcal{H}}$ the time-evolution operator, leading to 
	\begin{eqnarray}
		& g(T) =& \! \int \! \mathcal{D} \psi^L \mathcal{D} \psi^R  \bigg\langle \! \exp \Big\{i \int_0^T \dd t ( \mathcal{H}[\psi^R]- \mathcal{H}[\psi^L] )  \Big\}\! \bigg\rangle \nn \\
		& \qquad \times& \exp \bigg\{\!- \!\frac12 \sum_i\int_0^T \! \dd t  (  \psi^L_i \partial_t \psi^L_i+\psi^R_i \partial_t \psi^R_i) \bigg\} \, ,
		\label{eq:SFF_int}
	\end{eqnarray}
	where the denominator has been absorbed in the measure and  $\psi^R_i,\psi^L_i$ are the Grassmannian variables over the forward and backward time evolutions, respectively \footnote{The integration over the forward and backward contours resembles the Keldysh path integral, but with important differences in the time paths on which the action is defined, as explained in the supplementary material \cite{supp}}.
	
	After the Gaussian integration over the couplings, we can introduce collective fields in order to make the action quadratic in the fermions.
	In the usual SYK case, these are given by an averaged two-point function $G^{ab} (t, t^\prime)= \frac{1}{N}\sum_i\psi^a_i(t) \psi^b_i(t^\prime)$, with $a,b=L,R$. 
	In contrast, to account for the spatial profile of the power-law coupling, we introduce a local two-point function $G^{ab}_i (t, t^\prime)= \psi^a_i(t) \psi^b_i(t^\prime)$, similarly to \newred{\cite{Efetov:1997fw,Mirlin_2000_review,Xu_etal2020_sparseSYK}}.
	Imposing this definition at the path-integral level requires the introduction of a site-dependent auxiliary field $\Sigma_i^{ab}(t,t^\prime)$. Afterwards, we can perform two Gaussian integrals, one over the fermions and the other one over $G_{i}^{ab}$. 
	These manipulations, which are detailed in \cite{supp}, lead to   
	\begin{align}
		&g(T)= \int \mathcal{D} \Sigma_i \e^{-I[\Sigma_i]}\, ,  \label{eq:SFF_Sigma} \\
		&I= -\frac12 \sum_i   \Tr \log ( \partial -i \Sigma_i)\!+\!\sum_{abij} \frac{(-)^{a+b}}{4J^2}  \! \iint_0^T  A_{ij}^{-1} \Sigma_i^{ab} \Sigma_j^{ab} \nn ,
	\end{align}
	where the trace is over times and the $L,R$ indices and we defined $\partial_{at,bt^\prime} =\delta(t-t^\prime) \delta_{ab} \partial_{t^\prime}$. 
	
	{\it Saddle point equation.} Varying the effective action with respect to $\Sigma$ and assuming time-translation invariance, we get the saddle-point equation in frequency space
	\begin{equation}
		\frac{(-)^{a+b}}{J^2} \sum_j  A_{ij}^{-1}  \Sigma_j^{ba}(-\omega_n)= -(\omega_n \delta^{ab}- \Sigma_i^{ab}(\omega_n))^{-1} \, ,
		\label{eq:saddle_eq}
	\end{equation}
	where $\omega_n = (2 n +1) \pi / T $ are Matsubara frequencies  with $n \in \mathbb{Z}$.
	As discussed in \cite{supp}, this equation has a SU$(2)$ symmetry for each (positive) Matsubara frequency given by the adjoint action on the matrix indices of $\Sigma$, defining an infinite-dimensional symmetry group. 
	
	In principle, this equation may allow for several possible solutions. However, we will work in the ansatz where the dominating saddle does not depend on the site position. This assumption is consistent since averaging over the couplings restores space translation symmetry \footnote{In \cite{supp} we discuss further motivation for this assumption coming from an analysis of the eigenvalues of $A$: most of them vanish in the thermodynamic limit, leading to vanishing Fourier components of $\Sigma$, forcing the solution to be uniform.}. 
	As we will see, an initial decrease of the interaction range just produces subleading effects on the SFF, due to non-trivial cancellations among the eigenvalues of $A_{ij}$ in the one-loop corrections. 
		Deviations from this behaviour (see below) appear only for larger $\alpha$, when the PRBM localizes and non-uniform saddle-point solutions can emerge. 
By imposing  $\Sigma_i^{ab}= \tilde{\Sigma}^{ab}$, it is possible to obtain the same dominant saddles as in the $\alpha=0$ case \cite{supp}:
\begin{equation}
	\tilde\Sigma^{ab}(\omega_n) = 
	\begin{cases}
		\frac12 (\omega_n - \sqrt{\omega_n^2-4J^2}) \delta_{ab} & \omega_n > 2 J \, , \\
		\frac12 (\omega_n \delta_{ab} - i \sqrt{4 J^2-\omega_n^2} \sigma_3^{ab})   &0<\omega_n < 2J \, .
	\end{cases}
	\label{eq:saddle_sol}
\end{equation}

The solutions of Eq.~\eqref{eq:saddle_sol} for negative frequencies can be recovered using $\tilde\Sigma^{ab}(-\omega_n)=-\tilde\Sigma^{ba}(\omega_n)$.
One can check that the effective action evaluated on the homogeneous saddle point becomes linear in $N$ as in the usual SYK model, making the semiclassical approximation reliable.

Notably, for $|\omega_n|>2 J$, $\tilde\Sigma$ is proportional to the identity and therefore invariant with respect to the adjoint action of the SU$(2)$ symmetry discussed in \cite{supp}, while this is not the case for $|\omega_n|<2J$. In the latter regime, the classical solution spontaneously breaks SU$(2)$ to U$(1)$.  
The transformation in the coset space
SU(2)$/\text{U(1)}$ generates new equivalent solutions, leading to a contribution to the path integral proportional to the coset volume \footnote{This symmetry-breaking pattern resembles the one in SYK$_4$, where the ramp is generated by a spontaneous symmetry breaking of the U(1) time translation symmetry. Since our model is a free theory, we have a larger symmetry group that gets spontaneously broken.}.

{\it Quadratic fluctuations.} The above symmetry-breaking pattern implies zero modes, which are responsible for the leading correction to the classical action and the development of the ramp in the SFF. As noted in~\cite{Winer:2020mdc}, the ramp arises because, as $T$ increases, more Matsubara modes enter the symmetry-breaking regime $|\omega_n|<2J$, each bringing a contribution proportional to the coset volume. 
However, unlike~\cite{Winer:2020mdc}, the long-range decay can influence the number of zero modes.

To rigorously identify them, we analyse the Gaussian fluctuations around the saddle point solution
\begin{equation}
	\Sigma_i^{ab}(t,t^\prime) = \tilde\Sigma^{ab}(t,t^\prime)+\delta \Sigma_i^{ab}(t,t^\prime) \, .
\end{equation}
In contrast to the saddle itself, the fluctuations can have a spatial dependence. It is convenient to write $\delta \Sigma_{\red{i}}$ in frequency space and in a base that diagonalizes $A$. % the matrix $A_{ij}$. 
Denoting its eigenvalues as $\lambda_k$,the second-order action reads
%\vspace{-0.5cm}
%\sout{\begin{widetext}
		%\begin{equation}
		%g(T) =  e^{- I_\text{cl}}\int \mathcal{D} \delta\Sigma_i \exp \bigg\{\frac{1}{4 J^2 T^2} \sum_{k,a,b,\omega_1,\omega_2} \delta \Sigma^{ab}_k(\omega_1,\omega_2) K^{ab}_{k}(\omega_1,\omega_2) \red{\delta} \Sigma^{ba}_k(-\omega_2,-\omega_1) \bigg\} \, ,
		%\end{equation}
		%\end{widetext}}
		\red{\begin{equation}
				\delta I^{(2)} \! = \! \sum_{k,a,b \atop \omega_1,\omega_2} \! \delta \Sigma^{ab}_k(\omega_1,\omega_2) \frac{K^{ab}_{k}(\omega_1,\omega_2)}{4 J^2 T^2} \red{\delta} \Sigma^{ba}_k(\!-\!\omega_2,\!-\! \omega_1)
		\end{equation}}
		where $I_\text{cl}$ is the classical action contribution and
		\begin{equation}
			K^{ab}_{k}(\omega_1,\omega_2)=\frac{(-)^{a+b}}{\lambda_k}- \frac{1}{J^2} \tilde{\Sigma}^{aa}(\omega_1) \tilde{\Sigma}^{bb}(-\omega_2) \, .
			\label{eq:K_def}
		\end{equation}
		The zero modes arise from the condition $K^{ab}_{k}(\omega_1,\omega_2)=0$ and appear in the symmetry-breaking saddle for $a \neq b$ and $\omega_1=\omega_2$. %[as one can see inserting Eq.~\eqref{eq:saddle_sol} into Eq.~\eqref{eq:K_def}]. 
		In this case, Eq.~\eqref{eq:K_def} reduces to
		\begin{equation}
			K^{ab}_{k}(\omega,\omega)=-\lambda_k^{-1}+ 1 \, .
			\label{eq:zero_K}
		\end{equation}
		Therefore, the exponential ramp is generated by modes with eigenvalues $\lambda_k=1$. 
		
		{\it Eigenvalues of $A$.} $A$ can be diagonalized by going into Fourier space \cite{supp}. The constant vector is an eigenvector with eigenvalue $\lambda_{k=0}=1$, so we always have at least one zero mode. The other eigenvalues are \footnote{Since we are interested in the $N \to \infty$ limit, we do not notationally distinguish between even and odd $N$.}.
		\begin{equation}
			\lambda_k=\sum_{\Delta=1}^{N/2} \Delta^{-2 \alpha}  \cos \left(\frac{2 \pi}{N}\Delta k \right) \Big/ \sum_{\Delta=1}^{N/2} \Delta^{-2 \alpha}  \, .
			\label{eq:eigen_N_finite}
		\end{equation}
		The limit $N \to \infty$ must be taken with care since both numerator and denominator can diverge. 
		Analytic formulas can be derived for the cases when $k$ is comparable to $N$ or much smaller than $N$ \cite{supp}. 
		The eigenvalues corresponding to modes with $k\propto N$ are given by 
		\begin{equation}
			\label{eq:eigen_1}
			\lambda(\tilde{k}) = \begin{cases}
				0 & \alpha \le 1/2 \, , \\
				\frac{\text{Re} (\text{Li}_{2 \alpha}(e^{i \tilde{k}}))}{\zeta (2 \alpha)} & \alpha > 1/2 \, ,
			\end{cases}
		\end{equation}
		where Li is the polylogarithm and \sout{where} we defined $\tilde{k} = \frac{2 \pi}{N} k$, $-\pi<\tilde{k}<\pi$.
		The eigenvalues of the modes with $k=O(1)$, instead, are 
		\begin{equation}
			\lambda_k\! = \! 
			\begin{cases}
				(-1)^k \!+ \!\frac{\pi ^2 k^2 }{3-2 \alpha} \, _1F_2(\frac{3-2 \alpha}{2} ;\frac{3}{2},\frac{5-2 \alpha}{2} ;-\frac{ k^2 \pi ^2}{4}) & \alpha \le 1/2 \\
				1 & \alpha > 1/2 	,
			\end{cases}
			\label{eq:eigen_2}
		\end{equation}
		where $ _pF_q$ is the generalized hypergeometric function. 
		A comparison between the analytic eigenvalues in the thermodynamic limit and the ones computed numerically at finite $N$ can be found in Fig.~\ref{fig:eigen}.
		
		\begin{figure}
			\includegraphics[width=0.38\textwidth]{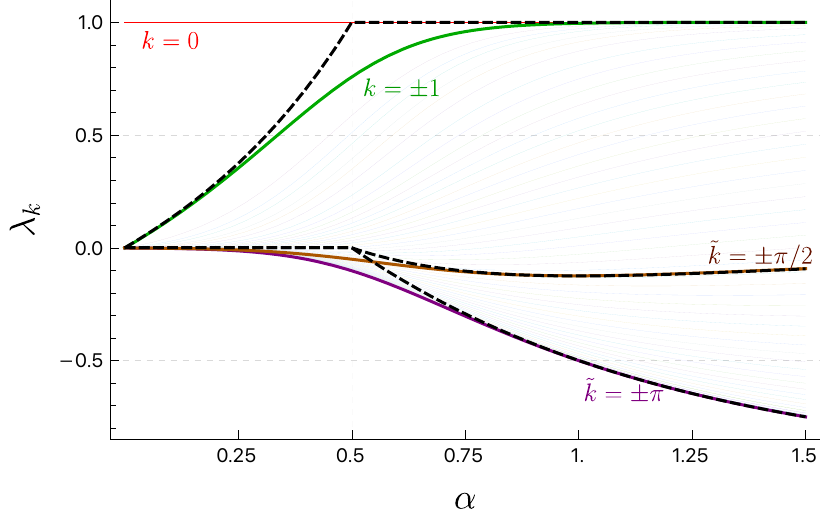}
			\caption{\justifying Eigenvalues $\lambda_k$ computed numerically for $N=1001$ (colored) and analytically at $N\to\infty$ (black dashed). A transition is signaled by the non-differentiable point at $\alpha=1/2$.}
			\label{fig:eigen}
		\end{figure}
		
		A drastic change happens at $\alpha=1/2$: as shown in Eq.~\eqref{eq:zero_K}, we have a zero mode whenever $\lambda_k=1$. The above solutions imply that for $\alpha>1/2$ an infinite number of zero modes should appear in the thermodynamic limit, each of them bringing a factor proportional to the coset volume. This should lead to a rapid increase of the ramp, if the saddle point we found is still dominating. However, as we argue in \cite{supp}, the same behaviour of the eigenvalues breaks down perturbation theory around the uniform saddle~\eqref{eq:saddle_sol}, signalling that we are entering a new regime where either the physics is described by another saddle or it becomes non-perturbative. Both imply a transition at $\alpha=1/2$; remarkably, this coincides with the critical point of the single-body PRBM. This transition has also been identified in other long-range models~\cite{Sahu:2021lgw} and coincides with the value of $\alpha$ at which the normalization $\mathcal{N}(\alpha)^2$ ceases to diverge in the thermodynamic limit. However, the value of $\alpha$ is not universal, and quantum spin systems with similar power-law interaction~\cite{Kastner_2011,Koffel_2012,Hauke:2013acj} can exhibit dynamical transitions at significantly different values.

		{\it Analytic SFF.} We now have all the ingredients to present the analytic expression for the SFF (see \cite{supp}).
		This is the product of three contributions: the classical action responsible for the slope, the time-symmetric fluctuations, which contain the zero modes and are responsible for the exponential ramp, and the massive fluctuations that are not time-symmetric and are responsible for the late-time corrections to the ramp. 
		The first is spatially homogeneous, while the two 1-loop contributions acquire a momentum dependence once $\alpha\neq 0$.
		
		For $JT$ large and for $\alpha<1/2$, the SFF simplifies to
		\begin{widetext}
			\begin{equation}
				\log \frac{g(T)}{2^N} = N \sum_{j=1}^\infty \left[ \frac{(-1)^{j+1}}{j} \frac{2 J_1(2JjT)}{2JjT} \right]+\frac{JT}{\pi} \log \left(\frac{16}{e^2}\frac{N}{JT}\right)+\frac{JT}{\pi} \sum_{k \neq 0} \bigg[ (1+ \lambda_k) \frac{\text{arctanh} \sqrt{\lambda_k}}{\sqrt{\lambda_k}}-1 +\log (1-\lambda_k)\bigg] \, ,
				\label{eq:g_analy}
			\end{equation}
		\end{widetext}
		where $J_1$ is a Bessel function of the first kind.
		Notably, for $\alpha \sim 0$ we have $\lambda_{k \neq 0} \sim N^{-1}$, the $\lambda_k$-dependent part cancels in a non-trivial way and becomes subleading for large $N$ recovering the result in \cite{Winer:2020mdc}. 
		A comparison between Eq.~\eqref{eq:g_analy} and numerical data for $\alpha = 0.3$ is presented in Fig.~\ref{fig:SFF}b, demonstrating excellent agreement. The $\alpha=0$ case would be indistinguishable from $\alpha=0.3$.  In general, for $\alpha < 1/2$, the eigenvalue contribution to \eqref{eq:g_analy} is subleading, as most eigenvalues vanish in the thermodynamic limit \eqref{eq:eigen_1}. The essentially unmodified form of the SFF highlights that, for $0\leq \alpha < 1/2$, the variable-range SYK$_2$ model is robust under connectivity reduction, consistent with the ergodic phase of the PRBM model.

		{\it SFF and localization.} In this section, we discuss the various regimes of the SFF for $\alpha \ge 1/2$. 
		The analytical analysis displays a breakdown of perturbation theory around the uniform saddle point at $\alpha=1/2$, at the same point where the single-particle PRBM displays an ergodic to non-ergodic transition. 
		For $\alpha>1/2$, the single-particle PRBM becomes gradually less delocalized, encounters an Anderson critical point at $\alpha = 1$, and enters an integrable regime at $\alpha > 3/2$ \cite{Mirlin_etal1996}. From a path integral point of view, it is reasonable to expect that, as the system localizes, it will break the space translation symmetry we assumed for the saddle-point solution in the small-$\alpha$ region. This is supported by the fact that at early times the SFF takes larger values (in particular at the dip time) than what is predicted analytically, indicating that there may be another saddle that dominates the classical action. This makes the dip height a good candidate for identifying the ergodic/non-ergodic transition. As it can be seen in  Fig.~\ref{fig:collapse}a, the dip starts to rise considerably in the vicinity of $\alpha\approx 0.5$. 
		A characteristic value of $\alpha$ where there is a qualitative change can be obtained similarly to the knee voltage in
		transistors. In this spirit, the intersection of a linear fit of the large-$\alpha$ part of the curve with the ordinate yields $\alpha =0.501 \pm 0.006$. 
		
		In the regime $1/2\leq \alpha \leq 3/2$, a secondary plateau emerges: it begins after the early-time oscillations and persists until it merges into the ramp characteristic of $\alpha < 1/2$ (see Fig.~\ref{fig:SFF}b).
		It becomes more prominent at larger system sizes \cite{supp} and signals a pre-thermalization regime~\cite{Baumgartner_etal2024}. The height of this secondary plateau grows with $\alpha$ until it reaches the height of the late-time one around $\alpha \sim 3/2$, as expected for an integrable theory. 
		To better quantify this transition, we define the logarithm of the ratio between the two plateau heights, 
		\begin{equation}
			h= \left\langle \frac{1}{N} \log \frac{2^{-N/2}}{g(T)} \right\rangle_T. 
			\label{eq:2plateau_distance}
		\end{equation}
		Here, $2^{-N/2}$ is the theoretical value of the late-time plateau, while the time average is taken over the interval $JT \in [10,15]$, where the secondary plateau has already developed. The remaining curvature in the SFF $g(T)$ is a small $N$ effect due to the proximity of the ramp. The value of $h$ at different $N$ is displayed in Fig.~\ref{fig:collapse}b.
		
		\begin{figure}
			\centering
			\includegraphics[width=0.49\textwidth]{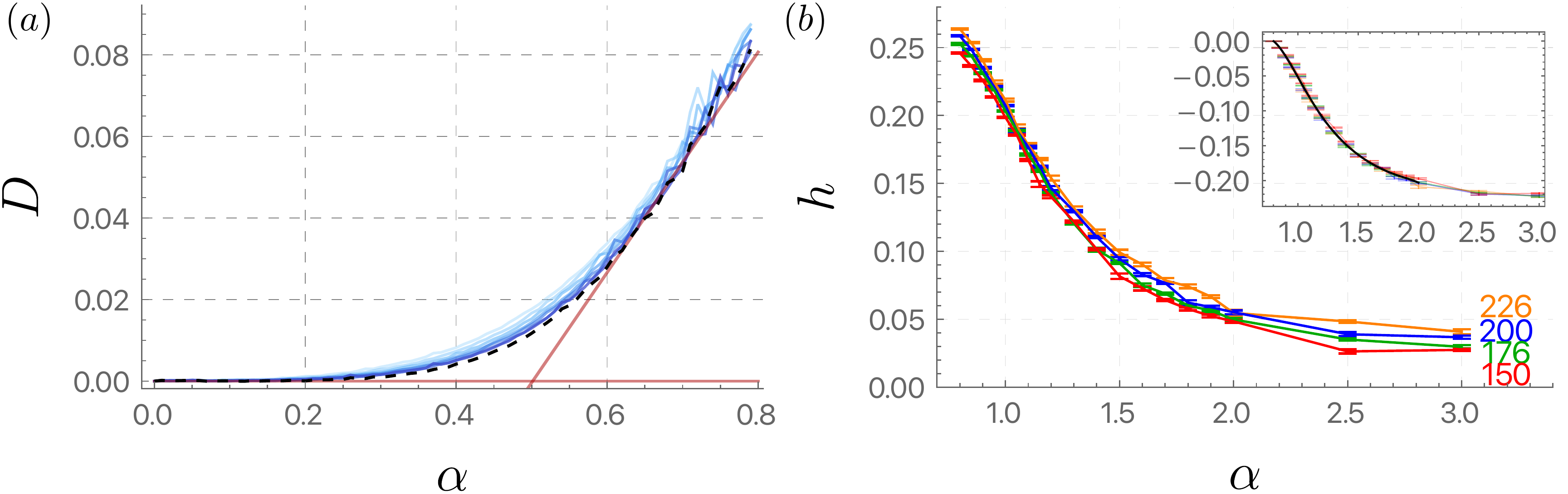}
			\caption{ \justifying 
				(a) Dip position as a function of $\alpha$ for different system sizes from $N=100$ (violet) to $400$ (light blue) in steps of $50$, averaged over $\sim 4 \red{\cdot 10^6}$ samples. The dip time is given by $T_\text{Dip} \sim 2.408$, and we define $D = \frac{1}{N} \log \frac{g(T_\text{Dip},\alpha)}{g(T_\text{Dip},\alpha=0)}$, effectively measuring the shift of the dip relative to the $\alpha=0$ SFF. The black dashed line shows an extrapolation to $N \to \infty$. 
				A linear fit to the curve at $\alpha\ge 0.6$ intersects $D=0$ at a characteristic value of $\alpha=0.501 \pm 0.006$.
				(b) Plateaus distance $h$,  defined in Eq.~\eqref{eq:2plateau_distance}, averaged over $\sim 4 \red{\cdot 10^7}$ samples for $N=226,200,176,150$. In the inset, the curves are vertically shifted to reveal a universal behavior. The black line represents a fourth-order polynomial fit.}
			\label{fig:collapse}
		\end{figure}

		By applying a constant vertical shift to $h$ to align the values at $\alpha=0$, the entire curves display an excellent collapse onto each other. A polynomial fourth-order fit of the collapsed curve above $\alpha=0.8$  reveals a change in concavity at $\alpha = 1.02$, in the vicinity of the point  where the PRBM has its Anderson transition. 
			This suggests that transitions within SYK$_2$-type models can be identified through features of the SFF such as the point where its plateau height grows more quickly upon a change of the system parameters.
		
		Within the regime accessible to our numerics, it is not possible to see a sharp transition at $\alpha=3/2$, corresponding in the PRBM to the onset of the integrable regime. We do, however, observe a qualitative disappearance of the ramp (Fig.~\ref{fig:SFF}b). 
		To understand whether this PRBM transition is imprinted onto the SFF of our model, it would be desirable to access larger system sizes, where the secondary plateau is more prominent. However, the SFF for SYK$_2$ has fluctuations that grow with $N^{T}$ \cite{legramandi-talwar}, making it prohibitively expensive to gather sufficient statistics for convergence, which in our case reaches the order of $10^7$ samples at $N = 226$. 
		The implications of these large fluctuations are discussed in \cite{supp}.

		{\it Conclusions.} We have discussed how the SFF captures distinct regimes in the variable-range SYK$_2$ model. 
		Key features include the dip height, 
		as an indicator of ergodic--non-ergodic transitions, and a secondary plateau, which appears at the onset of the non-ergodic regime, starts to grow more rapidly at a point suggestive of the position of Anderson criticality, and merges into the late-time plateau as the system localizes.
		
		These findings provide useful insights for the study of many-body localization (MBL), where numerics is often limited to small systems and where the existence of a transition in the thermodynamic limit is a matter of ongoing debate~\cite{Sierant_2025}. Our model offers a controlled setting in which many-body signatures of localization transitions can be directly connected to well-understood single-particle phenomena. For instance, the dip height has been used as a marker of transitions between holographic and non-holographic behavior in sparse SYK models \cite{Orman_Preskill2024}. Similarly, the emergence of a secondary plateau has been observed and studied \cite{Gharibyan_etal2018,Hopjan_Lev2023}. Since pre-ramp features have been linked to nearly conserved charges \cite{Baumgartner_etal2024}, it would be compelling to investigate whether the secondary plateau can be understood in terms of localized integrals of motion \cite{serbyn_etal2013,huse_etal2014}, further bridging our framework with core questions in MBL.

		Finally, our analysis can be extended to other models, such as sparse SYK. The adjacency matrix $M$ introduced in~\cite{Xu_etal2020_sparseSYK} shares key properties with the matrix $A$ in Eq.~\eqref{eq:A_def}, such as a uniform eigenvector with eigenvalue $1$. Exploiting the present framework may provide insights into the transitions appearing as the sparsification rate increases.

		{\it Acknowledgements.} We thank Gopal Chandra Santra, David Pascual Solis, Neil Talwar and Alex Windey for interesting discussions.
		This work was supported by the Swiss State Secretariat for Education, Research and lnnovation (SERI) under contract number UeMO19-5.1; the Ministry of University and Research through FARE grant for the project DAVNE (Grant R20PEX7Y3A), and through project DYNAMITE QUANTERA2\_00056, in the frame of  ERANET COFUND QuantERA II – 2021 call co-funded by the European Union (H2020, GA No 101017733); the European Union - Next Generation EU, Mission 4, Component 2 - CUP E53D23002240006; the Provincia Autonoma di Trento, and Q@TN, the joint lab between University of Trento, FBK—Fondazione Bruno Kessler, INFN—National Institute for Nuclear Physics, and CNR—National Research Council.
		S.B.\ acknowledges CINECA for the use of HPC resources under ISCRA-C projects ISSYK-2 (HP10CP8XXF)   and DISYK (HP10CGNZG9).

	{\it Data availability statement.} The datasets generated for figure 1 and 3 are available in the \href{https://doi.org/10.5281/zenodo.19359383}{Zenodo repository}.

		\bibliography{refs}
		
	\newpage
	\clearpage
	
	\begin{widetext}
		
		\renewcommand{\theequation}{S\arabic{equation}}
		\setcounter{equation}{0}
		\renewcommand{\thefigure}{S\arabic{figure}}
		\setcounter{figure}{0}
		\renewcommand{\thetable}{S\arabic{table}}
		\setcounter{table}{0}

		\setcounter{secnumdepth}{2}
		
		\renewcommand{\thesection}{\Alph{section}}

		{
			\let\clearpage\relax
			\let\newpage\relax\maketitle
		}

	\section*{Supplementary material}

		In this supplemental material, we provide detailed explanations and derivations to complement the calculations presented in the main text.  
		In Section~\ref{sec:alpha=0}, we show some intermediate steps in the derivation of Eq.~\eqref{eq:SFF_Sigma} and discuss how to recover the $\alpha=0$ case, corresponding to the quadratic SYK model discussed in Ref.~\cite{Winer:2020mdc}.  
		Section~\ref{sec:sym} contains a detailed analysis of the symmetries of the saddle-point equation and a discussion of its solutions.  
		In Section~\ref{subsection:eigenvalues}, we derive the explicit form of the eigenvalues of the matrix $A$ in the $N \to \infty$ limit.  
		Sections~\ref{sec:ramp} and \ref{sec:1-loop} focus on the calculation of the spectral form factor (SFF) up to the 1-loop level.  
		These contributions are particularly important at late times, where the classical action vanishes.  
		In Section~\ref{sec:ramp}, we analyze time-translation-invariant fluctuations, including the zero modes responsible for the ramp.  
		Section~\ref{sec:1-loop} incorporates the leading contribution from the soft modes.  
		By combining these contributions with the classical action, which dominates at early times, we derive in Section~\ref{eq:SFF_all} an analytic expression for the SFF, used in the main text for comparison with numerical results.  
		In Section~\ref{sec:perturbation}, we go beyond the 1-loop terms to explore perturbative corrections to the classical saddle points with respect to time-translation-invariant configurations.  
		This calculation demonstrates that the perturbative series becomes ill-defined for $\alpha > 1/2$.  
		Finally, in Section~\ref{sec:num}, we provide details on the numerical methods employed in the main text \red{and we discuss in Section~\ref{sec:SFF_PRBM}  the SFF for the PRBM model.}

		\subsection{Path integral for the spectral form factor and $\alpha=0$ case} \label{sec:alpha=0}
		
		In this section we show some details about the path integral calculation presented in the main text and explain how to consistently take the $\alpha \to 0$ limit in the hopping terms \eqref{eq:model_action} to recover the quadratic SYK model.
		The starting point of this section is the expression of the spectral form factor (SFF), Eq.~\eqref{eq:SFF_int} of the main text, which can be more explicitly written as
		\begin{equation}
			g(T) = \int \mathcal{D} \psi^L \mathcal{D} \psi^R \left\langle\exp \bigg\{-\sum_{i=1}^N\int_0^T \dd t \Big( \frac12 \psi^L \partial_t \psi^L+\frac12\psi^R \partial_t \psi^R +\sum_{j=i+1}^N \frac{J_{ij}}{\mathcal{N}}a_{ij} (\psi_i^L \psi_j^L-\psi_i^R \psi_j^R) \Big)  \bigg\} \right\rangle \, . \\
		\end{equation}
		\red{This expression is formally very similar to a Keldysh path integral, with a crucial difference: unlike the Keldysh path integral, where the forward and backward branches form a single connected contour, the SFF involves two disconnected contours, each with its own periodic boundary conditions due to the separate traces in $|Z(iT)|^2$. This disconnection implies that there is no explicit coupling between the contours, and the correlation between them are introduced by the ensemble average.}
		
		From here, it is straightforward to perform the Gaussian integral over the couplings, which leads to the following expression:
		\begin{equation}
			g(T) =\int \mathcal{D} \psi^L \mathcal{D} \psi^R \exp \bigg\{-\sum_{i=1}^N\int_0^T \dd t \frac12 \Big( \psi^L \partial_t \psi^L+\psi^R \partial_t \psi^R \Big) -\frac{J^2}{2\mathcal{N}^2}\sum_{j =i+1}^N a_{ij}^2 \Big(\int_0^T \dd t (\psi_i^L \psi_j^L-\psi_i^R \psi_j^R) \Big)^2 \bigg\} \, .
		\end{equation}
		
		To find the saddle-point solution, one typically introduces a two-point function field corresponding to the fermionic correlator \cite{MaldacenaStanford_2016}. 
		As mentioned in the body of the paper, differently from the usual SYK case, for our purposes this two-point function field cannot be spatially averaged, since local fields are needed to keep account of the spatial envelope. We thus use the identity
		\begin{equation}
			1=\prod_i\int \mathcal{D}G_i \delta \big(G^{ab}_i (t, t^\prime)- \psi^a_i(t) \psi^b_i(t^\prime)\big)   = \prod_i\int \mathcal{D}G_i \mathcal{D} \Sigma_i \exp \bigg\{\frac{i}{2}  \sum_{a,b} \iint_0^T \dd t \dd t^\prime \Sigma^{ab}_i(t,t^\prime) (G^{ab}_i (t, t^\prime)- \psi^a_i(t) \psi^b_i(t^\prime)) \bigg\} \,.
			\label{eq:coll_fields}
		\end{equation}
		Notice that the way we introduced $G$ and $\Sigma$ implies the following relations:
		\begin{equation}
			\label{eq:G_Sigma_symm}
			G^{ab}_i(t_1,t_2) = -  G^{ba}_i(t_2,t_1) = G^{ba}_i(t_2,t_1)^* \, , \qquad \Sigma^{ab}_i(t_1,t_2) = -  \Sigma^{ba}_i(t_2,t_1) = \Sigma^{ba}_i(t_2,t_1)^* \, .
		\end{equation}

		Plugging this expression into the SFF and expanding the square term, we can replace the fermions with $G_i$, yielding
		\begin{equation}
			\begin{split}
				g(T) &= \int \mathcal{D} \psi^a \mathcal{D}G_i \mathcal{D} \Sigma_i \exp \bigg\{-  \frac12 \sum_{i=1}^N \sum_{ab}\iint_0^T \dd t  \dd t^\prime \bigg[ \Big( \psi_i^a(t) (\delta(t-t^\prime) \delta_{ab} \partial_{t^\prime} -i \Sigma_i^{ab}(t,t^\prime))\psi_i^b(t^\prime)\Big)   \\
				& \qquad \qquad \qquad \qquad \quad-i \Sigma_i^{ab}(t,t^\prime) G_i^{ab}(t,t^\prime) +\frac{J^2}{2}   (-)^{a+b} \sum_{j \neq i}^N A_{ij} G_i^{ab}(t,t^\prime) G_j^{ab}(t,t^\prime) \bigg]\bigg\} \, ,
				\label{eq:integraL}
			\end{split}
		\end{equation}
		where we also used $A_{ij}=A_{ji}$.
		
		From here, it is now instructive to discuss how one can obtain the usual SYK$_2$ spectral from factor \cite{Winer:2020mdc} at $\alpha \to 0$. 
		The main difference so far in the $\alpha\to 0$ limit with respect to usual SYK$_2$, is how the propagators have been introduced in Eq.~\eqref{eq:coll_fields}, namely, as mentioned above, as a product of two Majorana spinors on a single site instead of averaging over all the positions. Another subtle point is that we are setting $i \neq j$ in the summation over $j$. These points are actually related: the standard SYK literature \cite{MaldacenaStanford_2016,Saad_etal2018} usually takes advantage of the fact that for the Grassmannian variables we have $\psi_j \psi_j = 0$, permitting one to extend the sum also to $j=i$. This would correspond in our case to setting $A_{ii}=1/\mathcal{N}^2$ instead of Eq.~\eqref{eq:A_def}.
		We have not done that, since we would have also needed to enforce the constraint $G_i(t,t_1)G_i(t,t_2)=0$. For the purpose of this section, it is useful to set $A_{ii}=1/\mathcal{N}^2$ in order to exactly recover the usual SYK case. The discussion that follows \sout{holds also} \red{also holds} for $A_{ii}=0$ up to $1/N$ corrections.

		By following the above convention, we now have $A_{ij}(\alpha=0)=1 /\mathcal{N}^2$ for all $i,j$, and therefore $A$ is a degenerate matrix with the single non-vanishing eigenvector
		\begin{equation}
			v_0 = \frac{1}{\sqrt{N}} (1 \dots 1)\,,
			\label{eq:v1_def}
		\end{equation}
		corresponding to the eigenvalue $\lambda_0 = 1$. 
		We call $v_k$ with $k \neq 0$ the basis vectors that span the remaining degenerate subspace of $A$. 
		To evaluate the path integral, it is convenient to use this basis, in which the propagator reads  
		\begin{equation}
			\hat{G}_0^{ab} = \frac{(G_1^{ab}, \dots , G_N^{ab}) \cdot v_0}{\sqrt{N}} = \frac{1}{N} \sum_i G_i^{ab} \, , \qquad \hat{G}_k^{ab} = (G_1^{ab} \dots G_N^{ab}) \cdot v_k\,.
		\end{equation}
		Here, we have added a factor $\sqrt{N}$ simply to make the notation consistent with the one usually used in the context of the standard SYK. 
		
		After these manipulations and using $\mathcal{N}(\alpha =0)^2=N$, the SFF reads
		\begin{equation}
			\begin{split}
				g(T) &= \int \mathcal{D} \psi^a \mathcal{D}G_i \mathcal{D} \Sigma_i \exp \bigg\{-  \frac12 \sum_{ab}\int_0^T \dd t  \dd t^\prime \bigg[ \sum_{i=1}^N \Big( \psi_i^a(t) (\delta(t-t^\prime) \delta_{ab} \partial_{t^\prime}-i \Sigma_i^{ab}(t,t^\prime))\psi_i^b(t^\prime)\Big)   \\
				& -i \sqrt{N}\hat\Sigma^{ab}_0(t,t^\prime) \hat{G}_0^{ab}(t,t^\prime) -i \sum_{k \neq 0}\hat\Sigma^{ab}_k(t,t^\prime) \hat{G}^{ab}_k(t,t^\prime) +\frac{J^2 N}{2}   (-)^{a+b} \hat{G}_0^{ab}(t,t^\prime)^2 \bigg]\bigg\} \, .
			\end{split}
		\end{equation}
		This expression is linear in the propagators $\hat{G}_k^{ab}$, allowing us to integrate them out. The result of this integration will produce some delta functions $\delta(\hat\Sigma_k)$. Since $\{v_k\}$ span an orthonormal basis, we must have $v_0 \cdot v_k = 0$ for all $k \neq 0$, meaning that the degenerate subspace of $A$ is spanned by vectors whose components sum to zero.
		This implies that these delta functions can be reorganized as 
		\begin{equation}
			\prod_{k \neq 0}\delta(\hat\Sigma_k)=\prod_{k \neq 0} \delta \left(\sum_i \Sigma^{ab}_i (v_k)_i \right)= N \prod_{i > 1} \delta\left(\Sigma^{ab}_1-\Sigma^{ab}_i \right) \, .
			\label{eq:Sigma_dirac}
		\end{equation}
		\red{This delta-functions restore spatial homogeneity of the action:} if we now call $\Sigma^{ab}=\hat\Sigma^{ab}_1$ and  $G^{ab}=\hat{G}^{ab}_0$, we exactly retrieve the familiar expression of the standard SYK$_2$ model \cite{Winer:2020mdc}, 
		\begin{equation}
			\begin{split}
				g(T) = \int \mathcal{D} \psi^a \mathcal{D} G \mathcal{D} \Sigma \exp \bigg\{-&  \frac12 \sum_{ab}\int_0^T \dd t  \dd t^\prime \bigg[ \sum_{i=1}^N \Big( \psi_i^a(t) (\delta(t-t^\prime) \delta_{ab} \partial_{t^\prime}-i\Sigma^{ab}(t,t^\prime))\psi_i^b(t^\prime)\Big) \\ 
				-&i N \Sigma^{ab}(t,t^\prime)G^{ab}(t,t^\prime) +\frac{J^2 N}{2}   (-)^{a+b} G^{ab}(t,t^\prime)^2 \bigg]\bigg\} \, .
			\end{split}
		\end{equation}

		Before concluding this section, let us briefly discuss how one can take the $\alpha \to 0$ limit after the integration of the fermions and the $G_i$, i.e., from Eq.~\eqref{eq:SFF_Sigma}.
		For doing so, it is important to impose a normalization for the path integral. Following Ref.~\cite{Winer:2020mdc}, we divide with respect to the same integral but with the Gaussian part only, i.e., without the $\Tr \log$ term. The idea is that the Gaussian part dominates in the interaction-less case $J \to 0$, and, since $JT$ is the only dimensionless quantity, this renormalization also imposes $g(0)=1$. 
		Moreover, this choice becomes important when we want to recover the $\alpha=0$ limit. Indeed, the matrix $A_{ij}^{-1}$ in this case is not defined due to infinite eigenvalues. However, thanks to the normalization, we have that the path integral is not diverging and the Gaussian terms with infinite eigenvalues approach Dirac delta functions as the ones in Eq.~\eqref{eq:Sigma_dirac}, giving the same results as in Ref.~\cite{Winer:2020mdc}.
		
		\subsection{Symmetries and solutions to the saddle point equation} \label{sec:sym}
		
		In this section, we discuss the symmetries and some manipulations of the saddle point equation \eqref{eq:saddle_eq}. First of all, it is convenient to re-write it in matrix form with respect to the replica indices $ab$, 
		\begin{equation}
			\sum_j  A_{ij}^{-1}  (\omega_n 1- \Sigma_i(\omega_n))\sigma_3 \Sigma_j^{t}(-\omega_n) \sigma_3= -J^2 \, .
		\end{equation}
		Second, Eq.~\eqref{eq:G_Sigma_symm} implies that $\Sigma_i(\omega_n)=\Sigma_i(\omega_n)^\dagger$.
		A natural way for a symmetry to act on $\Sigma$ is through the adjoint action $\Sigma_i (\omega_n)  \to U_i (\omega_n) \Sigma_i (\omega_n) U_i (\omega_n)^{-1}$. Since we want the transformed $\Sigma$ to be hermitian, we have to require that $U_i (\omega_n)$ is a representation of the special unitary group SU(2). 
		To have an actual symmetry, we must impose two more constraints, the first one is that $U_i(\omega_n)$ is site independent (for this reason from now on we will drop the subscript $i$). The second condition is a relation between the positive and negative frequencies, which transform in a different but correlated way, so that just one of them is really independent. If we fix $\omega_n >0$, we find the following transformation rules 
		\begin{equation}
			\Sigma_i (\pm\omega_n)  \to U (\pm\omega_n) \Sigma_i (\pm\omega_n) U (\pm\omega_n)^{-1} \, , \qquad U(- \omega_n) = \sigma_3 U(\omega_n)^* \sigma_3 \, .
		\end{equation}
		Notice that for each positive Matsubara frequency we have an independent SU(2) transformation, leading to an effectively infinite-dimensional symmetry group.

		Now, we will discuss how to get the saddle point solution Eq.~\eqref{eq:saddle_sol}. In order to make progress with Eq.~\eqref{eq:saddle_eq}, we use the conditions of Eq.~\eqref{eq:G_Sigma_symm}, which implies $\Sigma_i^{ba}(- \omega_n) = -\Sigma_i^{ab}(\omega_n)$, to write Eq.~\eqref{eq:saddle_eq} in terms of just one frequency:
		\begin{equation}
			\frac{1}{J^2} \sum_j  A_{ij}^{-1}  \Sigma_j^{ab}(\omega_n)=(-)^{a+b}(\omega_n \delta^{ab}- \Sigma_i^{ab}(\omega_n))^{-1} \, .
			\label{eq:saddle_eq2}
		\end{equation}
		Now, we perform a Fourier transform of the equation above along the spatial direction, so that we can diagonalize the matrix $A_{ij}^{-1}$, as shown in the next section. The result of this operation is
		\begin{equation}
			\hat\Sigma_{-k}^{ab}(\omega_n)= \lambda_k J^2 (-)^{a+b} \reallywidehat{(\omega_n \delta^{ab}- \Sigma^{ab}(\omega_n))^{-1}}_k,
		\end{equation}
		where $\lambda_k$ are the eigenvalues of $A$, while the hat indicates the Fourier transform.
		From the exact expression of the eigenvalues given in section \ref{subsection:eigenvalues}, we known that, for $\alpha \sim 0$ we have $\lambda_0=1$ and $\lambda_{k \neq 0} \sim 0$, which implies that $\Sigma_{k \neq 0} \sim 0$. Therefore, for $\alpha \sim 0$ we have a single non-vanishing solution, which is independent of the site position. In the main text, we have assumed that this saddle dominates at least up to $\alpha = 1/2$. This assumption is justified by the fact that for $\alpha$ small most of the eigenvalues are very close to zero. However, for $\alpha>1/2$ the situation changes, and we have that most of the eigenvalues are non-zero.
		
		By setting  $\Sigma_i^{ab}= \tilde{\Sigma}^{ab}$ and noticing that $\sum_j  A_{ij}^{-1}=\mathcal{N}^{-2}$ (as consequence of the fact that $v_1$ is an eigenvalue also for $A^{-1}$ with eigenvector $\lambda_1^{-1}= \mathcal{N}^{-2}$), we get
		\begin{equation}
			\frac{1}{J^2} \tilde\Sigma^{ab}(\omega_n)= (-)^{a+b}(\omega_n \delta^{ab}- \tilde\Sigma^{ab}(\omega_n))^{-1} \, .
			\label{eq:saddle_reduced}
		\end{equation}
		This equation can be easily solved and the dominant saddle is given in Eq.~\eqref{eq:saddle_sol}.
		
		\subsection{More details on the eigenvalues in the thermodynamic limit} \label{subsection:eigenvalues}
		
		In this section, we give some more details on the derivation of the eigenvalues of $A$, for arbitrary $\alpha$.
		For calculating them, it is convenient to perform a Fourier transform and go to momentum space. In the following, we will suppress the time and $L,R$ indices in order to make the notation lighter and we will assume that $N$ is odd, which will not affect the result in the large-$N$ limit:
		\begin{equation}
			\begin{split}
				\sum_{i,j}^N A_{ij} G_i G_j &= \frac{1}{ N} \sum_{k,p} \sum_{ij} A_{ij} e^{i \frac{2 \pi}{N}(ik+jp)} \hat{G}_k \hat{G}_p \\
				&= \frac{1}{N \mathcal{N}} \sum_{k,p} \sum_{i=1}^N \sum_{\Delta=-\lfloor N/2 \rfloor}^{\lfloor N/2 \rfloor}  \frac{e^{i \frac{2 \pi}{N}(i(k+p))} e^{i \frac{2 \pi}{N}\Delta k}}{|\Delta|^{2 \alpha}} \hat{G}_k \hat{G}_p  \\
				&= \sum_{k} \left(\frac{1}{\mathcal{N}}\sum_{\Delta=-\lfloor N/2 \rfloor}^{\lfloor N/2 \rfloor}  \frac{ e^{i \frac{2 \pi}{N}\Delta k}}{|\Delta|^{2 \alpha}} \right)\hat{G}_k \hat{G}_{-k}\, , 
			\end{split}
			\label{eq:F_trans_eigen}
		\end{equation}
		where we have introduced in the second step the distance between two sites $\Delta = i-j$, the floor function $\lfloor \bullet \rfloor$ and we defined the Fourier transform as 
		\begin{equation}
			\hat{G}_k = \frac{1}{\sqrt{N}} \sum_{k=-\lfloor N/2 \rfloor}^{\lfloor N/2 \rfloor}  e^{i \frac{2 \pi}{N}ik} G_i \, .
		\end{equation}
		With this, the eigenvalues of $A_{ij}$ can be read from the term in the parenthesis of Eq.~\eqref{eq:F_trans_eigen}, and are
		\begin{equation}
			\lambda_k= \frac{1}{\mathcal{N}}\sum_{\Delta=-\lfloor N/2 \rfloor}^{\lfloor N/2 \rfloor}  \frac{ e^{i \frac{2 \pi}{N}\Delta k}}{|\Delta|^{2 \alpha}} =\frac{ \sum_{\Delta=1}^{\lfloor N/2 \rfloor} \Delta^{-2 \alpha}  \cos \left(\frac{2 \pi}{N}\Delta k \right) }{ \sum_{\Delta=1}^{\lfloor N/2 \rfloor} \Delta^{-2 \alpha} } \, .
		\end{equation}
		as in Eq.~\eqref{eq:eigen_N_finite}. In principle, since $A$ is a symmetric real matrix, we could have \sout{diagonalize} \red{diagonalized} it also by using an orthogonal change of \sout{coordinate} \red{coordinates}. This is defined by:
		\begin{equation}
			\label{eq:real_fourier}
			G_j = \sqrt\frac{1}{N} \hat{G}_0  + \sqrt\frac{2}{N} \sum_{k>0} \left[ \cos \left( \frac{2 \pi}{N}ik  \right) \hat{G}_k - \sin \left( \frac{2 \pi}{N}ik  \right) \hat{G}_{-k}  \right] \, .
		\end{equation}
		
		The next step is to take the thermodynamic limit. For $\alpha < 1/2$, $\mathcal{N}$ is divergent while the rest of the expression is finite, which leads to a vanishing contribution. For $\alpha > 1/2$ $\mathcal{N}$ becomes convergent can now define $\tilde{k} = \frac{2 \pi}{N} k$ with $-\pi<\tilde{k}<\pi$ to get the result in \eqref{eq:eigen_1}.
		
		Now, let us give some more details about the derivation of Eq.~\eqref{eq:eigen_2}. The starting point is the finite-$N$ expression \eqref{eq:eigen_N_finite}, from which we proceed by expanding the cosine in a power series:
		\begin{equation}
			\begin{split}
				\lambda_k&=\frac{ \sum_{\Delta=1}^{\lfloor N/2 \rfloor} \Delta^{-2 \alpha}  \cos \left(\frac{2 \pi}{N}\Delta k \right) }{ \sum_{\Delta=1}^{\lfloor N/2 \rfloor} \Delta^{-2 \alpha} } \\
				&=\frac{\sum_{n=0}^\infty   \frac{(-1)^n}{2n!}\left(\frac{2 \pi}{N} k \right)^{2n} \sum_{\Delta=1}^{\lfloor N/2 \rfloor} \Delta^{2n-2 \alpha}  }{\sum_{\Delta=1}^{\lfloor N/2 \rfloor} \Delta^{-2 \alpha} } \\
				&=1+  \sum_{n=1}^\infty   \frac{(-1)^n}{2n!}\left(2 \pi k \right)^{2n} \frac{1}{N^{2n}} \frac{  \sum_{\Delta=1}^{\lfloor N/2 \rfloor} \Delta^{2n-2 \alpha}  }{ \sum_{\Delta=1}^{\lfloor N/2 \rfloor} \Delta^{-2 \alpha} } \, .
			\end{split}
		\end{equation}
		For finite $k$, it is possible to take the $N \to \infty$ limit of the term in the last term alone,
		\begin{equation}
			\lim_{N \to \infty} \frac{1}{N^{2n}} \frac{ \sum_{\Delta=1}^{\lfloor N/2 \rfloor} \Delta^{2n-2 \alpha}  }{ \sum_{\Delta=1}^{\lfloor N/2 \rfloor} \Delta^{-2 \alpha} } =
			\begin{cases}
				4^{-n}\frac{1-2 \alpha }{1-2(\alpha-n)} & \alpha \le 1/2 \\
				0 & \alpha > 1/2 .
			\end{cases}
		\end{equation}
		By plugging this expression back into the summation, one obtains the result in Eq.~\eqref{eq:eigen_2}. Notice that all the $\lambda_k$, as a function of $\alpha$, are continuous, but they are not smooth at $\alpha = 1/2$.

		\subsection{Analytic calculation of the ramp}
		\label{sec:ramp}
		
		In this section, we calculate the leading contribution to the path integral which is responsible for the exponential ramp. We slightly change our conventions in order to follow the notation of the Supplemental Material of Ref.~\cite{Winer:2020mdc}, so that the interested reader to better compare our results to the ones in Ref.~\cite{Winer:2020mdc}, which can be recovered in the $\alpha \to 0$ limit. The same notation will be used in section \ref{sec:perturbation}.
		This convention uses a slightly different but equivalent Hamiltonian for the $R$ action, namely 
		\begin{equation}
			\mathcal{H}_R = -\frac{i}{\mathcal{N}} \sum_{i<j}^N J_{ij }a_{ij} \psi_i^R \psi_j^R \, ,
		\end{equation}
		which allows us to get rid of the $(-)^{a+b}$ term in Eq.~\eqref{eq:integraL}. Moreover, we also absorb an $i$ in the definition of $\Sigma$, which sends $\Sigma \to -i \Sigma$. After these notational changes, the SFF reads 
		\begin{equation}
			g(T) = \int \red{\left( \prod_i \mathcal{D} \Sigma_i \right)} \exp \bigg\{\frac12 \sum_i  \Tr \log (\delta(t-t^\prime) \delta_{ab} \partial_{t^\prime} -\Sigma_i^{ab}(t,t^\prime))   +\frac{\mathcal{N}^2}{4J^2} \sum_{ab\red{,i}j} \int_0^T \dd t  \dd t^\prime   A_{ij}^{-1} \Sigma_i^{ab}(t,t^\prime) \Sigma_j^{ab}(t,t^\prime)\bigg\} \, .
		\end{equation}
		
		To analyze the leading contribution to the exponential ramp, we assume time-translation invariance, expressed as $\Sigma_i^{ab}(t, t^\prime) = \Sigma_i^{ab}(t - t^\prime)$. This assumption effectively captures the semiclassical saddle point and the zero modes relevant to the ramp. However, it does not account for the massive modes that govern the late-time behavior of the ramp and the eventual transition to the plateau, which require a separate discussion (see section \ref{sec:1-loop}).
		The time-invariance symmetry makes it convenient to Fourier transform into frequency space. We also define the dimensionless quantities $(\sigma_n^i)^{ab} = \Sigma^{ab}_i(\omega_n)/J$ and $x_n = \omega_n/J$, where $\sigma_n^i$ is a Hermitian two-dimensional matrix. The SFF for the time-translation invariant (t.t.i.) configurations can be expressed as an infinite product of finite-dimensional integrals, 
		\begin{eqnarray}
			& &g^{\text{t.t.i.}}(T) = \prod_{n>0} g_n(T) \, \qquad \qquad  g_n(T) = \frac{\int \red{\prod_i} \dd \sigma_n^i e^{I[\sigma_n]}}{\pi^{2N} \det(A/\mathcal{N})^2} \, , \label{eq:SFF_rampappendix} \\[2mm]
			& &I[\sigma_n] = \sum_i \frac{1}{2} \Tr \log \left[\left(1-i \frac{\sigma_n^i}{x_n}\right)\left(1-i \frac{\sigma_n^i}{x_n}\right)^\mathcal{\red{T}} \, \right] -\frac{\mathcal{N}^2}{2} \sum_{i,j} A_{ij}^{-1} \Tr(\sigma^i_n \sigma^j_n) \, ,
		\end{eqnarray}
		where we used $\sigma_{-n}^i=(\sigma_n^i)^\mathcal{\red{T}}$ for multiplying the positive and negative contributions together and thus \sout{restrict} \red{restricting} the product to $n>0$. Each integral is already normalized with the same normalization discussed in section \ref{sec:alpha=0}.
		
		The dominant saddle points are the same as those given by Eq.~\eqref{eq:saddle_sol}, up to the change of conventions:
		\begin{equation}
			\label{eq:saddle_solution_new-conventions}
			\tilde\sigma_n = 
			\begin{cases}
				-\frac{i}{2} (x_n - \sqrt{x_n^2-4})  & x_n> 2 \, , \\
				-\frac{i}2 (x_n - i \sqrt{4-x_n^2} \sigma_3)   &x_n < 2 \, . 
			\end{cases}
		\end{equation}
		The saddle point action can be obtained by evaluating $I[\tilde\sigma_n]$ and has been explicitly computed in Ref.~\cite{Winer:2020mdc}: 
		\begin{equation}
			I[\tilde\sigma_n] = 
			\begin{cases}
				N \left[\log \left(\frac{1}{2} \left(1+ \sqrt{1-\frac{4}{x_n^2}}\right)-\frac{1}{x_n^2}\right)+\frac{x_n}{4} (1-\sqrt{1-\frac{4}{x_n^2}})^2\right] & x_n > 2 \, ,\\[2mm]
				N \left[-\log x_n^2 + \frac12 (x_n^2 - 2)\right] & x_n < 2 \, .
			\end{cases}
		\end{equation}

		To include the contribution beyond the saddle point approximation, we have to expand the fields in the fluctuations around it 
		\begin{equation}
			\sigma_n = \tilde{\sigma}_n + \delta \sigma_n \, , \qquad I[\tilde\sigma_n+\delta\sigma_n]=I[\tilde\sigma_n]+\delta I_n[\delta \sigma_n] \, ,
		\end{equation}
		where we chose a basis for the fluctuation $\delta \sigma_n$ that already diagonalizes the matrix $A$, so that the the quadratic action reads
		\begin{equation}
			\label{eq:second order action}
			\delta I_n^{(2)} = -\frac12 \sum_k \Tr \left(\tilde\sigma_n \delta\sigma_n^k\tilde{\sigma}_n \delta \sigma_n^k\right)-\frac{1}{\lambda_k} \Tr(\delta \sigma^k_n  \delta\sigma^k_n) \, .
		\end{equation} 
		A convenient way for performing the integral over $\delta \sigma_n^k$ is to expand it on the Pauli matrix basis, 
		\begin{equation}
			\delta \sigma_n^k = \sum_\mu y^k_\mu \sigma_\mu\,, 
		\end{equation}
		where $\sigma_4 = 1$. In these new coordinates
		\begin{equation}
			\delta I_n^{(2)} = -\sum_k \Big( \tilde\sigma_n^{LL}\tilde\sigma_n^{RR}((y^{k}_1)^2+(y^{k}_2)^2) + \frac{\tilde\sigma_n^{RR}}{2}((y^{k}_3)^2-(y^{k}_2)^4)+ \frac{\tilde\sigma_n^{LL}}{2}((y^{k}_3)^2+(y^{k}_2)^4)+\frac{1}{\lambda_i} ((y^{k}_1)^2+(y^{k}_2)^2+(y^{k}_3)^2+(y^{k}_4)^2) \Big) \, .
			\label{eq:delta_I_vacuum}
		\end{equation}
		
		Due to the form of the saddle point solution \eqref{eq:saddle_solution_new-conventions}, it is convenient to split the calculation in two pieces, depending on weather $x_n \gtrless 2$. For $x_n<2$ it is possible to explicitly notice that we have zero modes corresponding and $k=0$, which are due to the fact that $\lambda_0= 1$ and $\tilde\sigma_n^{LL}\tilde\sigma_n^{RR}=-1$. A naive integration of \eqref{eq:delta_I_vacuum} over $y^0_{1,2}$ would therefore lead to a divergent result, since the variables seem to run over the real axis. However, these zero-modes are generated by the degeneracy of the symmetry broke saddle-point solution, which means that \sout{$y^0_{1,2}$ can only explore the null directions, which, as discussed in \cite{kamenev_mezard1999}, are restricted to the saddle point manifold SU(2)$/\text{U(1)}$.}\red{, as discussed in \cite{kamenev_mezard1999}, $y^0_{1,2}(\theta,\phi)$ can only explore the vacuum manifold SU(2)$/\text{U(1)}$ which is parameterized by the coordinate $\theta,\phi$.}  This means that the integral over the Goldstone modes just contributes as the vacuum-manifold volume:
		\begin{equation}
			\label{eq:0_volume}
			\red{\int \dd \theta \dd \phi} = 4 \pi N (1-x_n^2/4)\, .
		\end{equation}
		Let us now break down this expression. 
		The vacuum manifold is given by \sout{a} the coset space SU$(2)/$U$(1)$, which is a two-dimensional sphere. The volume of a two-dimensional sphere with radius $r$ is given by $4 \pi r^2$, so, we just need to understand what is $r$ in our case.
		First of all, we point out that by rotating the fluctuations in such a way that $\delta \sigma_i$ diagonalizes $A$, we are going into the real basis defined in \eqref{eq:real_fourier}. 
		The eigenvector corresponding to $\lambda_0=1$ is 
		\begin{equation}
			\sigma^0_n = \frac{1}{\sqrt{N}} \sum_i \sigma^i_n \, .
		\end{equation}
		On the saddle point solution, we have that $ \sigma^i_n = \tilde{\sigma}_n$, which means $\sigma_n^0 = \sqrt{N} \tilde\sigma_n$. 
		Now, the radius of the vacuum manifold is given by the coefficient of the symmetry-breaking part of the saddle solution, which is the one proportional to $\sigma_3$ in Eq.~\eqref{eq:saddle_solution_new-conventions}, so that we have $r = \sqrt{ N (1-x_n^2/4)}$, where we already include the $\sqrt{N}$ dependence discussed before.  This exactly reproduces the result in Eq.~\eqref{eq:0_volume}
		
		The integral over the other coordinates $y_{3,4}^k$ in Eq.~\eqref{eq:delta_I_vacuum} is a simple Gaussian integration. The overall result for $x_n <2$ is:
		\begin{equation}
			g_{n} = e^{I[\tilde{\sigma_n}]} N \sqrt{4-x_n^2} \prod_{k \neq 0} \frac{1}{(1-\lambda_k)\sqrt{1+\lambda_k(1-x_n^2+\lambda_k)}} \, .
			\label{eq:gn_smallx}
		\end{equation}
		This completes the calculation of the time-\sout{trnaslation}\red{translation} invariant part of the SFF up to the one-loop term for $x_n<2$.
		As a sanity check of Eq.~\eqref{eq:gn_smallx}, we can recover the $\alpha =0$ case: from Eq.~\eqref{eq:gn_smallx} it is very simple to take the $\alpha \to 0$ limit, since, thanks to the path integral renormalization, the eigenvalues $\lambda_k$ appear in a regular way in the final result. In the large-$N$ limit, this gives the same result as what we would have gotten directly from Ref.~\cite{Winer:2020mdc}, since the product in Eq.~\eqref{eq:gn_smallx} collapses to 1 in this limit.
		
		Now, we can analyze the one-loop calculation for $x_n>2$.
		In this case, we do not have zero modes and we can simply perform the Gaussian integral of the quadratic action in Eq.~\eqref{eq:delta_I_vacuum}. The contributions to the SFF in this regime is given by
		\begin{equation}
			g_{n} = e^{I[\tilde{\sigma_n}]} \prod_{k} \left(1+\frac{\lambda_k}{2}(2-x_n^2+x_n \sqrt{x_n^2-4}) \right)^{-2} \, .
			\label{eq:gn_bigx}
		\end{equation}
		
		The final expression of the one-loop contribution to the SFF for the time-translation invariant configurations as defined in Eq.~\eqref{eq:SFF_rampappendix} is given by the product of the factors given by Eq.~\eqref{eq:gn_smallx} and Eq.~\eqref{eq:gn_bigx} over all the Matsubara frequencies.
		This expression can be explicitly evaluated at late time $JT \gg 1$. Neglecting the classical action terms (which give a vanishing contribution for \red{large $JT$}), we can substitute the sum over the Matsubara modes by integrals, 
		\begin{equation}
			\begin{split}
				\log g^{\text{t.t.i.}}(T) = - I_\text{ramp}&=\frac{JT}{2 \pi} \bigg[\int_0^2 \left(\log N\sqrt{4-x^2} +\sum_{k \neq 0} \log  \frac{1}{(1-\lambda_k)\sqrt{1+\lambda_k(1-x^2+\lambda_k)}}\right) \dd x \\[2mm]
				& +\int_2^\infty  \sum_{k} \log \frac{4}{ \left(2+\lambda_k(2-x^2+x \sqrt{x^2-4}) \right)^2}  \dd x\bigg] \\[2mm]
				&=\frac{JT}{\pi} \bigg[\log (64 N /\e^3)+\sum_{k \neq 0} (1+ \lambda_k) \frac{\text{arctanh} \sqrt{\lambda_k}}{\sqrt{\lambda_k}}-1 \bigg]  \, . \\
			\end{split}
			\label{eq:ramp_cont}
		\end{equation}
		The name $I_\text{ramp}$ is due to the fact that this term is the one responsible for the exponential \sout{growing} \red{growth} of the ramp, as it can be seen from the fact that $I_\text{ramp} \propto JT$.
		
		Notice that for $\alpha \sim 0$, which implies $\lambda_k \sim 0$ for $k \neq 0$, the sum gives a negligible $O(1)$ contribution and we are left with just the first term. This result is the same as the one in Ref.~\cite[eq. (S29)]{Winer:2020mdc}, up to the corrections of some terms due to the fact that the vacuum volume and the one-loop contributions in that reference were not properly included.
		
		\subsection{Soft modes contribution}
		\label{sec:1-loop}
		
		In section \ref{sec:ramp} we have calculated the zero-mode volume and the one-loop determinant of the time-translation-invariant fluctuations. However, as shown in Eq.~\eqref{eq:K_def}, there are also contributions to the one-loop determinant that are not time-translation invariant. These are given by
		\begin{equation}
			\det{\tilde{K}^{ab}_{k}(\omega_1,\omega_2)}^{-\frac{1}{2}} \qquad \omega_1 \neq \omega_2 \, ,
		\end{equation}
		where
		\begin{equation}
			\tilde{K}^{ab}_{k}(\omega_1,\omega_2)=(-)^{a+b}- \frac{\lambda_k}{J^2} \tilde{\Sigma}^{aa}(\omega_1) \tilde{\Sigma}^{bb}(-\omega_2) \, 
		\end{equation}
		is the renormalized kernel, where we have already accounted for the correct path integral normalization discussed in section \ref{sec:alpha=0}. We can massage the expression for the determinant of $\tilde{K}$ as \sout{following} \red{follows}
		\begin{equation}
			- \frac{1}{2}\log \det  \tilde{K} = - \frac12 \log \Tr \tilde{K} = - \frac12 \sum_{a,b,k} \sum_{\omega_1 \neq \omega_2} \log \left( (-)^{a+b}- \frac{\lambda_k}{J^2} \tilde{\Sigma}^{aa}(\omega_1) \tilde{\Sigma}^{bb}(-\omega_2) \right) \, .
			\label{eq:log_1_loop}
		\end{equation}
		
		Further, we restrict the analysis to the soft modes, which are responsible for the leading contribution. The soft modes can be thought of as small deviations from the zero modes discussed around Eq. \eqref{eq:eigen_N_finite}, so they are determined by $a \neq b$ and $|\omega_{1,2}| < 2J$, but now the two frequencies differ by a small quantity $\omega_2 = \omega_1 + \epsilon$ with $|\epsilon| = 2 \pi |j| / T \ll 2J$. This last condition is met at late times, when $JT \gg 1$. Expanding $\tilde{K}$ in the small $\epsilon$ limit, we get 
		\begin{equation}
			- I_\text{soft} =  - \frac12 \sum_{-2J<\omega_1<2J} \sum_{k} \sum_{j \neq 0} \log \left(1+ \lambda_k \Big( -1 - i \frac{ \text{sign}(j) \epsilon }{\sqrt{4 J^2-\omega_1^2}} \Big) \right)=  -  \sum_{0<\omega_1<2J} \sum_k \sum_{j=1}^M \log \left((1- \lambda_k)^2 + \frac{ \epsilon^2 \lambda_k^2 }{4 J^2-\omega_1^2}  \right) \, .
		\end{equation}
		Here, $M$ is a positive integer counting the number of soft modes, so it must satisfy the relation $M \ll JT$. 
		
		The presence of $M$ results in an ambiguous expression, as there is no natural choice for its value. This ambiguity can be resolved by using the analytical properties of the kernel $\tilde{K}$ and its behaviour at infinity, which are the same as in the $\alpha=0$ case. Following \cite{Winer:2020mdc}, the first step is to notice that 
		\begin{equation}
			\int \dd \omega_1 \dd \omega_2 \log \tilde{K}^{ab}_{k}(\omega_1,\omega_2) = 0 \, .
		\end{equation}
		The idea now is to add this integral to regulate the determinant of the kernel and restrict it to the soft-mode contribution, which becomes
		\begin{equation}
			- I_\text{soft} = -  \sum_{0<\omega_1<2J,k} \left(\sum_{j=1}^M- \int_0^{M+\frac{1}{2}} \dd j\right) \log \left((1- \lambda_k)^2 + \frac{ \epsilon^2 \lambda_k^2 }{4 J^2-\omega_1^2} \Big) \right) \, .
		\end{equation}
		The limits of integration have been chosen in such a way to eliminate the $M$ dependence. In order to see this, we have to separate in the sum $\lambda_0=1$ from the rest of the eigenvalues, i.e., $I_\text{soft}=I_{\text{soft},{k=0}}+ I_{\text{soft},{k\neq 0}}$. For $k=0$, we have
		\begin{equation}
			- I_{\text{soft},{k=0}}=- \sum_{0<\omega_1<2J} 2 \left(\sum_{j=1}^M- \int_0^{M+\frac{1}{2}} \dd j \right) \log \left( \frac{2 j \pi/ T}{\sqrt{4 J^2-\omega_1^2}}  \right) \sim - \sum_{\omega_1>0}\log \left(T \sqrt{4 J^2-\omega_1^2} \right) \, ,
		\end{equation}
		where we have used Stirling's approximation for large $M$. Performing the sum over $\omega_1 = \pi (2n+1)/T \sim 2 \pi n/T $, we get
		\begin{equation}
			- I_{\text{soft},{k=0}}=- \frac{1}{2}\sum_{n< JT/\pi} \log \left[(2 \pi)^2 \left(\frac{J^2 T^2}{\pi^2}- n^2\right)\right] = \frac{JT}{\pi} \log \left(\frac{e}{4 JT}\right)
		\end{equation}
		where we have used again  Stirling's approximation. We can evaluate in a similar way the $k \neq 0$ sum:
		\begin{equation}
			- I_{\text{soft},{k\neq 0}}=- \sum_{0<\omega_1<2J}  \sum_{k \neq 0}\left(\sum_{j=1}^M- \int_0^{M+\frac{1}{2}} \dd j \right) \log \left((1- \lambda_k)^2 + \frac{ ( 2 \pi k \lambda_k /T)^2 }{4 J^2-\omega_1^2} \Big) \right) \sim \frac{1}{2} \sum_{0<\omega_1<2J}  \sum_{k \neq 0} \log (1-\lambda_k)^2 ,
		\end{equation}
		and the sum over $\omega_1$ trivially leads to
		\begin{equation}
			- I_{\text{soft},{k\neq 0}}=\frac{JT}{\pi} \sum_{k \neq 0} \log (1-\lambda_k) \, .
		\end{equation}
		Summing up these two contributions, we have \red{that} the 1-loop determinant \eqref{eq:log_1_loop} can be approximated by the soft modes contribution given by
		\begin{equation}
			-I_\text{soft} = - I_{\text{soft},{k= 0}}- I_{\text{soft},{k\neq 0}}=\frac{JT}{\pi} \log \left(\frac{e}{4 JT}\right)+\frac{JT}{\pi} \sum_{k \neq 0} \log (1-\lambda_k) \, ,
		\end{equation}
		which provides the leading correction to the exponential ramp calculated in section \ref{sec:ramp}.
		
		\subsection{Analytic expression for the SFF}
		\label{eq:SFF_all}
		
		In this section we will explain how to derive Eq.~\eqref{eq:g_analy}. Our calculation of the SFF $g(T)$ can be thought as divided in two contributions: one determined by the classical action $I_{\text{cl}}$, which is relevant at early times, while the second one by the 1-loop determinant. In the previous sections, we split this last contribution into two pieces, depending on whether they contained the zero modes (section \ref{sec:ramp}) or not (section \ref{sec:1-loop}). 
		Overall, the SFF is given by
		\begin{equation}
			g(T)= e^{-I_\text{cl}} e^{-I_\text{ramp}} e^{-I_\text{soft}} \, 
		\end{equation}
		where the last two contributions are the one computed in the previous sections, while for evaluating $I_\text{cl}$ we need to consider the action on the saddle point solution. Since the saddle point solution \eqref{eq:saddle_sol} is \sout{a} site-independent, the classical action reduces to
		\begin{equation}
			-I[\tilde{\Sigma}]= \frac12 \sum_i   \Tr \log ( \partial -i \tilde{\Sigma}) -\sum_{abj} \frac{(-)^{a+b}}{4J^2}  \! \iint_0^T  A_{ij}^{-1} \tilde\Sigma^{ab} \tilde\Sigma^{ab} =  \frac{N}{2} \left[  \Tr \log ( \partial -i \tilde{\Sigma}) -\sum_{ab} \frac{(-)^{a+b}}{2J^2}  \! \iint_0^T  \tilde\Sigma^{ab} \tilde\Sigma^{ab} \right]\,,
		\end{equation}
		which is the same action as in the $\alpha=0$ case. This means that we can borrow the classical action contribution from the one calculated in Ref.~\cite{Winer:2020mdc}, which reads
		\begin{equation}
			I_\text{cl}=N \log 2+N \sum_k \frac{(-1)^k}{k} \frac{2 J_1(2JkT)}{2JkT} \, ,
		\end{equation}
		where $J_1$ is a  is a Bessel function of the first kind.
		Combining all three terms, we arrive at Eq.~\eqref{eq:g_analy}.

		\subsection{Perturbative calculation}
		\label{sec:perturbation}
		
		This section aims to extend the one-loop calculation presented in section~\ref{sec:ramp} by setting up a perturbative calculation to determine the regimes in which the saddle point solution in Eq.~\eqref{eq:saddle_sol} can be trusted. In the $\alpha = 0$ case, we anticipate a well-defined perturbative series in $1/N$ as for the quadratic SYK model, however we need an understanding also of the $\alpha>0$ case. In this section, we will use the conventions discussed in section \ref{sec:ramp} and we will restrict our analysis to the time-symmetric contribution to the SFF.

		To evaluate perturbations around the saddle point solution $\tilde{\sigma}_n$, we need to expand the action $I[\sigma_n] = I[\tilde\sigma_n + \delta \sigma_n] = I[\tilde\sigma_n]+\delta I_n[\delta \sigma_n]$ to higher orders with respect to the quadratic one considered in Eq.~\eqref{eq:second order action}. Differently from what we have done in section \ref{sec:ramp}, we will expand the fluctuations on the Fourier basis:
		\begin{equation}
			\label{eq:per_action}
			\delta I_n[\delta \sigma_n] = -\frac12 \sum_i \sum_{m \ge 2} \frac{(-1)^m}{m} \Tr \left[ \left(\tilde\sigma_n \sum_{k} \frac{e^{\frac{2\pi i}{N} j k}}{\sqrt{N}}\delta\sigma_n^{k}\right)^m \right] - \sum_{k}\frac{1}{\lambda_k} \Tr(\delta\sigma^k_n \delta\sigma^{-k}_n) \, ,
		\end{equation}
		where now $\delta\sigma^{-k}_n=(\delta\sigma^{k}_n)^\dagger$. 
		We now separate the quadratic part of $\delta I_n$, which we name $\delta I^{(2)}_n$, from the higher orders, and then expand the exponential of the higher-order terms in a power series.
		At each order of this expansion, the integral is Gaussian and can therefore be performed explicitly, provided that we can determine the propagator, which is needed for evaluating the Wick contractions. 
		
		The quadratic action reads
		\begin{equation}
			\delta I^{(2)}[\delta \sigma_n] = - \frac12 \sum_{k,l\ge0} \sum_{abcd} (\delta \sigma_n^{-k})_{ab} (\delta \sigma_n^l)_{cd} K_{ab,cd}^{k,l} \,.
		\end{equation}
		Here we defined the inverse propagator
		\begin{equation}
			K_{ab,cd}^{k,l} = \delta_{ac} \delta_{bd} \delta_{kl} \left((\tilde{\sigma}_n)_{aa} (\tilde{\sigma}_n)_{bb} + \lambda^{-1}_k\right) \, ,
		\end{equation}
		which is nothing but Eq.~\eqref{eq:K_def} in the new conventions and restricting to the time-translation invariant case. Notice that for $x_n < 2$ we have spontaneously broken symmetries, which requires to remove the zero modes that correspond to $k=0$ before inverting $K$.
		Taken care of this subtlety, $K$ is diagonal in the indices $(k,a,b)$, and it can be easily inverted to give the propagator:
		\begin{equation}
			(K^{-1})_{ab,cd}^{k,l} = \delta_{ac} \delta_{bd} \delta_{kl} \frac{\lambda_k}{(\tilde{\sigma}_n)_{aa} (\tilde{\sigma}_n)_{bb} \lambda_k+1} .
			\label{eq:propagator}
		\end{equation}
		
		Excluding the classical action and 1-loop contribution, which do not play a role in the discussion, a generic perturbative expansion is defined by
		\begin{equation}
			\langle e^{\delta I_n - \delta I^{(2)}_n} \rangle = \sum_\alpha \frac{1}{\alpha !} \langle (\delta I_n - \delta I^{(2)}_n)^\alpha \rangle
		\end{equation}
		where now $\langle \bullet \rangle$ indicates the normalized Gaussian integration with measure $\delta I_n^{(2)}$. 
		The non-quadratic terms $\delta I_n - \delta I^{(2)}_n$ can be seen as interaction vertices and are given by
		\begin{equation}
			\delta I_n - \delta I^{(2)}_n=-\frac12 \sum_i \sum_{m \ge 3} \frac{(-1)^m}{m} \Tr \left[ \left(\tilde\sigma_n \sum_{k} \frac{e^{\frac{2\pi i}{N} jk}}{\sqrt{N}}\delta\sigma_n^{k}\right)^m \right] \, .
			\label{eq:pert_vertices}
		\end{equation}
		Since an odd number of $\delta \sigma_n$ \sout{insertion} \red{insertions} would result in a vanishing contribution, the first non-trivial term $C$ in the perturbative expansion is given by setting $\alpha = 1$ and $m=4$. More explicitly this term reads:
		\begin{equation}
			\begin{split}
				C=& - \frac{1}{8} \sum_{j} \sum_{k_1 \dots k_4} \sum_{a_1 \dots a_4} \frac{e^{\frac{2\pi i}{N} j(k_1+\dots+k_4)}}{N^2} (\tilde{\sigma}_n)_{a_1 a_1} \dots (\tilde{\sigma}_n)_{a_4 a_4} \langle (\delta \sigma_n)_{a_1 a_2}^{k_1}  (\delta \sigma_n)_{a_2 a_3}^{k_2}  (\delta \sigma_n)_{a_3 a_4}^{k_3}  (\delta \sigma_n)_{a_4 a_1}^{k_4} \rangle \\
				=& - \frac38 \sum_{k_1,k_2} \frac{1}{N} \left[\sum_a \frac{\lambda_{k_1} (\tilde{\sigma}_n)_{aa}^2}{(\tilde{\sigma}_n)_{aa}^2 \lambda_{k_1}+1} \cdot  \frac{\lambda_{k_2} (\tilde{\sigma}_n)_{aa}^2}{(\tilde{\sigma}_n)_{aa}^2 \lambda_{k_2}+1}  \right]\,, 
				%=& - \frac38 \sum_{k_1,k_2} \frac{\delta_{(k_1 + k_2) , 0}+\delta_{(k_1 + k_2 -N/2) , 0}}{N} \left[\sum_a \frac{\lambda_{k_1} (\tilde{\sigma}_n)_{aa}^2}{(\tilde{\sigma}_n)_{aa}^2 \lambda_{k_1}+1} \cdot  \frac{\lambda_{k_2} (\tilde{\sigma}_n)_{aa}^2}{(\tilde{\sigma}_n)_{aa}^2 \lambda_{k_2}+1}  \right]\ .
			\end{split}
		\end{equation}
		Due to the presence of the eigenvalues $\lambda_k$, this expression is hard to evaluate explicitly, but we can extract the large-$N$ behaviour in the relevant cases. 
		Let's start by considering the case where $\alpha = 0$. The only eigenvalue which does not scale with $1/N$ is $\lambda_0=1$, while we have $\lambda_{k \neq 0} = 1/N$. Since $\tilde{\sigma}$ is order $O(1)$,we have
		\begin{equation}
			C = O(1/N) \, , \qquad \alpha = 0\, .
		\end{equation}
		We expect this to still \sout{ne} \red{be} true for $\alpha \gtrsim 0$, since just a small number of eigenvalues with small $k$ are $O(1)$, as discussed in section \ref{subsection:eigenvalues}. However, for $\alpha > 1/2$ most of the eigenvalues become $O(1)$, giving the following contribution to the first term in the perturbative expansion
		\begin{equation}
			C = O(N) \, , \qquad \alpha >1/2\, .
		\end{equation}
		We expect the transition $C=O(1)$ to happen around $\alpha=1/2$, in correspondence to the kink of $\lambda_k$, even if a more in-depth study would be needed to confirm this. This signals the breakdown of the saddle-point approximation for $\alpha > 1/2$, which can be due either to entering a non-perturbative regime or to the appearance of a new saddle point.
		
		\begin{figure}
			\centering
			\includegraphics[width=1.00\textwidth]{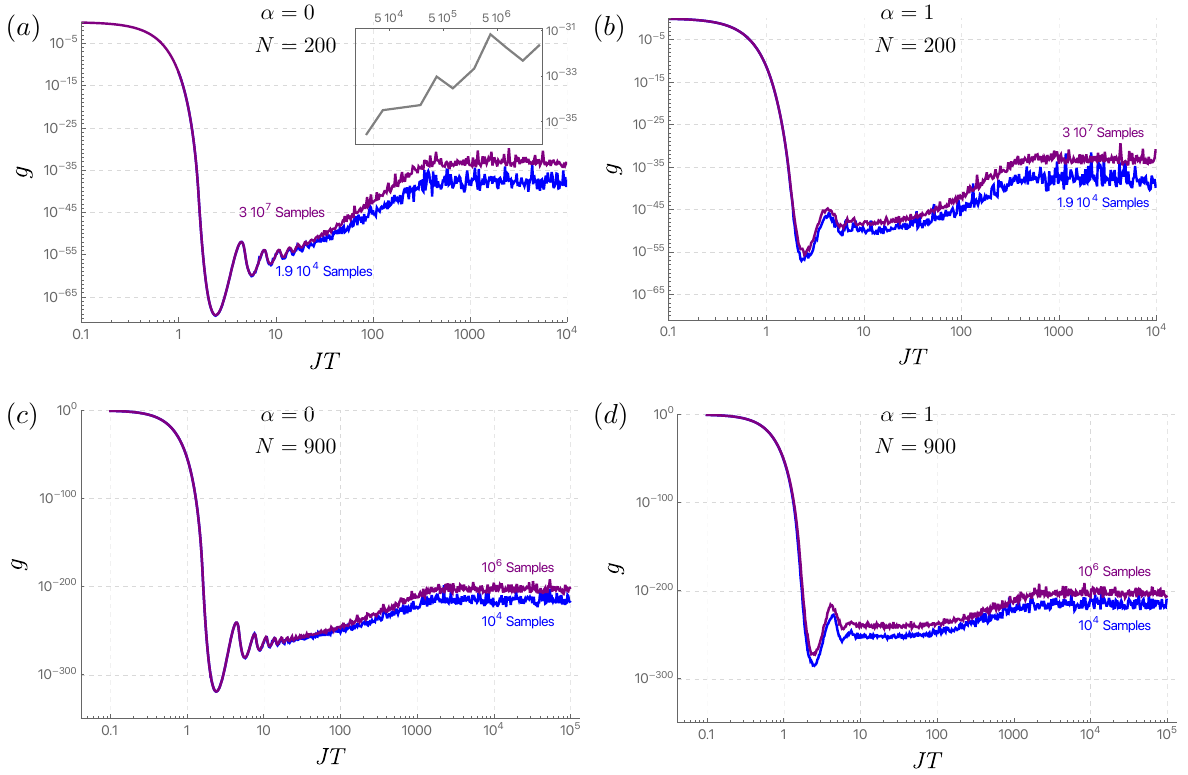}
			\caption{ \justifying 
				\red{(a)-(b) SFF for $N = 200$ and $\alpha = 0, 1$, respectively, averaged over $19 \cdot 10^3$ samples (blue) and $3 \cdot 10^7$ samples (violet). The offset between the two curves is due to insufficient averaging in the smaller sample size, which leads to a systematic upward bias. For $\alpha = 0$, the gap between the ramp and plateau increases over time and eventually saturates, while for $\alpha = 1$, the offset becomes visible at earlier times. The inset shows the averaged plateau value (computed over both time and disorder realizations) as a function of the number of samples. As the sample size increases, the plateau converges toward the theoretical value $2^{-N/2} = 8 \cdot 10^{-31}$. 
					(c)-(d) Similar plots for $N = 900$, with averages over $10^4$ samples (blue) and $10^6$ samples (violet). For $\alpha = 1$, the presence of the secondary plateau is stable and remains visible regardless of the sample size.}}
			\label{fig:complem}
		\end{figure}
		
		Notice that, in principle, we could have considered terms with $m>4$ in Eq.~\eqref{eq:pert_vertices}. 
		For $\alpha = 0$ this would have \sout{lead} \red{led} to subleading contributions \red{with} respect to $m=4$ due to the higher factors of $1/\sqrt{N}$ coming from the Fourier coefficient in equation \eqref{eq:pert_vertices}.
		For $\alpha > 1/2$ we don't need to consider more contributions because if just one term becomes non-perturbative, the whole series is.

		\subsection{Details on the numerics}
		\label{sec:num} 
		
		For the numerical analysis, the SFF has been reduced to the single particle eigenvalues $\mu_i$ of the hopping matrix $a_{ij} J_{ij}$ similarly to what is has been done e.g. in \cite{Liao_etal2020} for complex fermions:
		\begin{equation}
			g(T) =  \prod_{\mu_i >0} \cos^2 \left( \frac{\mu_i T}{2} \right) \, .
		\end{equation}
		While this expression allows to accelerate considerably the numerical simulation and explore quite large system sizes compared to SYK$_{q>2}$ models, the quadratic SYK exhibits extremely large fluctuations in the ramp regime even for $\alpha=0$. These fluctuations scale as $(N/T)^T$ \cite{legramandi-talwar}, presenting a challenge for accurately averaging the SFF in systems with large $N$.
		
		Insufficient statistical averaging leads to noticeable artifacts in the SFF, as illustrated in Fig. \ref{fig:complem}. These include not only large fluctuations but also a systematic downward shift of the SFF \red{after the dip}. This effect is particularly prominent in the late-time plateau, which stabilizes below the theoretical value of $2^{-N/2}$. This shift is also at the origin of the displacements between the various curves in \ref{fig:collapse}, which, however, can be corrected by a constant shift, as shown in the inset.
		
		\begin{figure}
			\centering
			\includegraphics[width=0.48\textwidth]{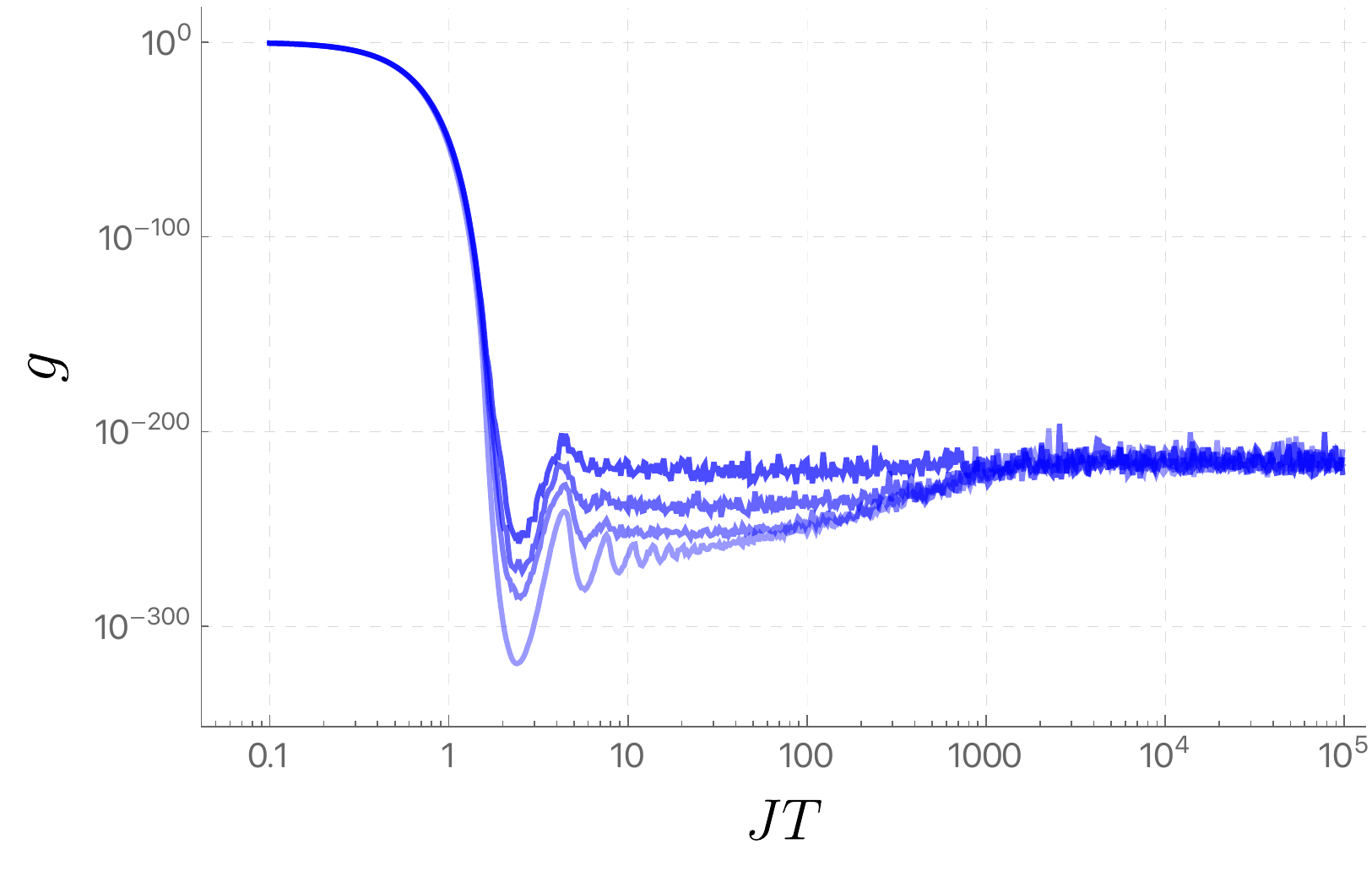}
			\caption{ \justifying The SFF for $N=900$ and 50$K$ samples at $\alpha=0,0.9,1.1,1.5$ (from lighter to darker shades of blue). Compared to the $N=200$ case shown in Figure~\ref{fig:SFF}b, the secondary plateau is significantly longer and flatter. }
			\label{fig:large_SFF}
		\end{figure}
		
		This limitation restricts us to relatively small sizes compared to those achievable with exact diagonalization of the hopping matrix, leaving \sout{to} the secondary plateau less time to develop. An example of the SFF for $N=900$ is shown in Fig. \ref{fig:large_SFF}. Although the statistical sample size is insufficient, as evidenced by the compression of the SFF after the slope, it is still apparent that the secondary plateau is significantly longer and flatter compared to that in Fig.~\ref{fig:SFF}. \red{This behavior can be understood from the fact that, in the quadratic SYK$_2$ model, the dip time is essentially independent of $N$, while the late-time plateau begins at $JT = 2N$, as shown in \cite{Liao_etal2020}. As a result, increasing $N$ creates a wider temporal window in which a secondary plateau can develop. This secondary plateau emerges once the initial oscillations of the SFF following the dip are damped (around $JT \sim 10$), and it persists until it intersects with the exponential ramp characteristic of the $\alpha = 0$ case.}

		\subsection{SFF for a PRBM model}
		\label{sec:SFF_PRBM} 
		
		\red{In this section, we present some numerical study of the SFF for a PRBM model. A similar analysis was previously carried out in~\cite{Hopjan_Lev2023}. Here, however, we adapt the single-particle Hamiltonian to the presence of Majorana fermions and define
			\begin{equation}
				\label{eq:PRBM_H}
				[\mathcal{H}_{\text{s.p.}}]_{ij} = \frac{i}{\mathcal{N}}  J_{ij} \, a(i-j) \, , 
			\end{equation}
			where $\mathcal{N}$, $J_{ij}$, and $a(i-j)$ are defined as in the main text.
			Note that if we had considered complex fermions instead of Majorana fermions, the $J_{ij}$ could have been taken from the Gaussian Unitary Ensemble (GUE), as done in~\cite{Hopjan_Lev2023}. }
		
		\red{We present numerical results for system sizes $N = 200$ and $N = 4000$, at different values of $\alpha$, in Fig \ref{fig:PRBM}. The results show that the different phases manifest also in the single-particle Hamiltonian, even if with different features. For $\alpha < 1/2$, the SFF is indistinguishable from the $\alpha = 0$ case, as one expects from the ergodic phase.
			As $\alpha$ increases the SFF does not display a secondary plateau (as for variable-range SYK) but it develops a pronounced dip that becomes longer and higher as $\alpha$ increases. This dip also flattens with increasing system size, as can also be noticed in \cite{Hopjan_Lev2023}, where system size $N=20000$ has been considered. A similar trend appears in the variable-range SYK model, where large system sizes are required for the secondary plateau to fully emerge and flatten. 
			The dip terminates when it encounters the ramp corresponding to $\alpha = 0$, which is linear instead of exponential as in the many-body system. The ramp disappears for $\alpha \ge 3/2$ when the dip reaches the plateau high, signaling the presence of an integrable phase. The onset of the plateau occurs at the Heisenberg time $t_{\rm H} = 2N$, just as in the variable-range SYK model. However, the interpretation of the Heisenberg time differs significantly between the two models: in the PRBM case, $N$ corresponds to the Hilbert space dimension, whereas in the SYK model, it denotes the number of Majorana fermions.}

		\begin{figure}
			\centering
			\includegraphics[width=1.00\textwidth]{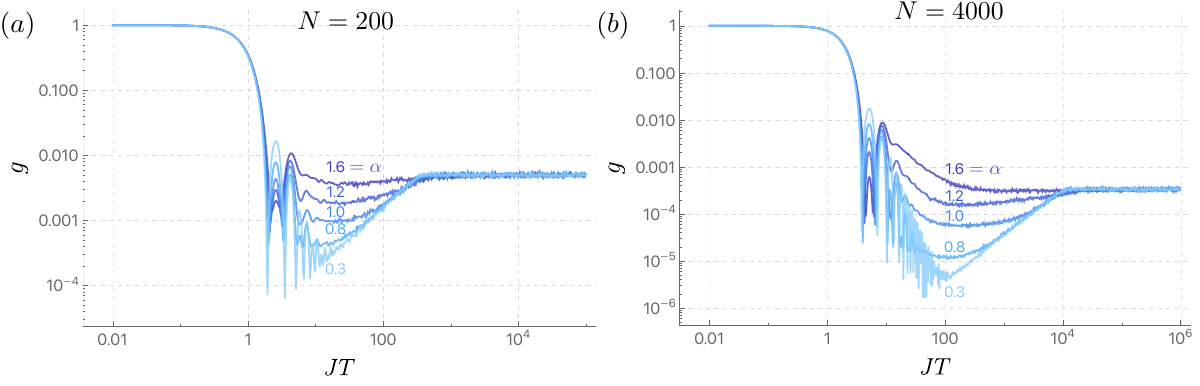}
			\caption{ \justifying 
				\red{The SFF for the PRBM model in Eq.~\eqref{eq:PRBM_H} for $N = 200$ (a) and $N=4000$ (b) averaged over $1.1 \cdot 10^3$ realizations at different values of $\alpha$. The transitions at different values of $\alpha$ are visible in the single-particle picture, even if they manifest themselves in a different way compared to the variable-range SYK$_2$ model. For $\alpha < 1/2$ the SFF does not significantly change (just $\alpha = 0.3$ is presented, the other ones would overlap with it). For $\alpha > 1/2$ the dip start growing and becoming longer and flatter, until it disappears in the late-time plateau for $\alpha > 3/2$.}}
			\label{fig:PRBM}
		\end{figure}

	\end{widetext}
		
	\end{document}